\documentclass[seceq,supplement]{ptptex}

\usepackage{amsmath}
\usepackage{graphicx}

\newcommand{\bv}[1]{{\boldsymbol #1}}

\title{%        %You can use \\ for explicit line-break
Behavior of pressure and viscosity at high densities for 
two-dimensional hard and soft granular materials
}

\author{%       %Use \scshape  for the family name
Michio Otsuki$^1$ \footnote{E-mail : otsuki@phys.aoyama.ac.jp}, 
Hisao Hayakawa$^2$ \footnote{ E-mail: hisao@yukawa.kyoto-u.ac.jp }
and Stefan Luding$^3$ \footnote{E-mail : s.luding@utwente.nl}
}

\inst{%         %Affiliation, neglected when [addenda] or [errata]
\small
${}^1$Department of Physics and Mathematics, 
Aoyama Gakuin University,
5-10-1 Fuchinobe, Sagamihara, Kanagawa, 229-8558, Japan, \\
${}^2$Yukawa Institute for Theoretical Physics, Kyoto University,  
Kitashirakawaoiwake-cho, Sakyo-ku, Kyoto 606-8502, Japan, \\
${}^3$ Multi Scale Mechanics, Faculty of Engineering Technology, 
       University of Twente, P.O.\ Box 217, 7500 AE Enschede, 
       Netherlands 
}

\abst{%         %this abstract is neglected when [addenda] or [errata]
The pressure and viscosity in two-dimensional
sheared granular assemblies are investigated numerically
for varying disks' toughness, degree of polydispersity and 
coefficient of normal restitution.
In the rigid, elastic limit of monodisperse systems, 
the viscosity is approximately inverse proportional to the area fraction 
difference from  $\phi_{\eta} \simeq 0.7$, 
but the pressure is still finite at $\phi_{\eta}$.
On the other hand, in moderately soft, dissipative and polydisperse systems, 
we confirm the recent theoretical 
prediction that both scaled pressure (divided by the kinetic temperature $T$)
and scaled viscosity (divided by $\sqrt{T}$) diverge at the same density, 
i.e., the jamming transition point $\phi_J > \phi_\eta$, %XXX
with the critical exponents $-2$ and $-3$, respectively. 
Furthermore, we observe that the critical region of the jamming transition 
disappears as the restitution coefficient approaches 
unity, i.e.\ for vanishing dissipation.

In order to understand the conflict between these two different 
predictions on the divergence of the pressure and viscosity,
the transition from soft to near-rigid particles is studied in detail
and the dimensionless control parameters are defined as ratios
of various time-scales. We introduce a dimensionless number,
i.e. the ratio of dissipation rate and shear rate,
that can identify the crossover from the scaling of very hard, 
i.e. rigid disks, in the collisional regime, to the scaling in the 
soft, jamming regime with multiple contacts.
}

\begin{document}

\maketitle

\section{Introduction}

One of the reasons for the growing interest in granular materials,
i.e. collections of interacting macroscopic particles
\cite{aronson,pouliquen,silbert,lutsko01,alam03,GDRMiDi,lutsko04,namiko05,kumaran,orpe07,hisao07a,namiko07,hisao07b,hayakawa08,kudrolli,Hatano08,Kumaran09,Hayakawa09s,Otsuki09s,Otsuki09r,Otsuki09,Otsuki:Jstat,Otsuki:EPJE,Luding09}
is the fact that these materials are different from ordinary matter \cite{jaeger}.
%The growing interest in granular materials,
%which are collections of dissipative macroscopic 
%particles \cite{aronson,pouliquen,silbert,lutsko01,alam03,GDRMiDi,lutsko04,namiko05,kumaran,orpe07,hisao07a,namiko07,hisao07b,hayakawa08,kudrolli,Hatano08,Kumaran09,Hayakawa09s,Otsuki09s,Otsuki09r,Otsuki09,Otsuki:Jstat,Otsuki:EPJE,Luding09},
%is due to the fact that granular materials are different from
%usual materials \cite{jaeger}. 
%Nevertheless, we believe that moderately dense
%and nearly elastic granular flows can be described
%by the hydrodynamic equations derived from kinetic theory
The pertinent differences do not preclude a description of (up to)
moderately dense and nearly elastic granular flows by hydrodynamic
equations with constitutive relations derived using kinetic theory
\cite{lutsko04,hisao07a,Brilliantov,JR85a, Dufty, Santos, Garzo, Soto, Brey, Goldhirsch03, Lutsko05}.
%This is also valid when correlations are involved, which are understood
%by fluctuating hydrodynamics theories associated with kinetic theory
When nontrivial correlations, such as long-time tails 
and long-range correlations, are present, one can apply
fluctuating hydrodynamic descriptions to granular fluids and the latter can
be obtained from kinetic theory as well.
\cite{kumaran,hisao07b,kudrolli,Kumaran09,Hayakawa09s,Otsuki09s,Otsuki09r,Otsuki09,Otsuki:Jstat,Otsuki:EPJE}.

Similar analysis cannot be applied to systems near the jamming transition.
Indeed, we know many examples when the behavior of very dense flows
cannot be understood by Boltzamnn-Enskog theory 
\cite{Ishiwata,Garcia,namiko07,Khain07,Khain09,Luding09,Otsuki:PTP, Otsuki:PRE}
due to effects like ordering or crystallization, excluded volume, anisotropy and
higher order correlations.  Therefore, to understand the rheology of
dense granular flows, such as the frictional flow \cite{silbert},
and the jamming transition itself \cite{Liu},
an alternative approach is called for.
%we need to adopt an alternative approach.

Recently, Otsuki and Hayakawa have proposed a mean-field theory to describe the 
scaling behavior close to the jamming transition \cite{Otsuki:PTP,Otsuki:PRE}
at density (area fraction) $\phi_J$.
They predicted that both pressure and viscosity are proportional 
to $(\phi_J - \phi)^{-4}$. Therefore, the scaled pressure, divided by the 
kinetic granular temperature ${T} \propto (\phi_J - \phi)^{-2}$, 
is proportional to $(\phi_J-\phi)^{-2}$, while the scaled viscosity, 
divided by $\sqrt{T} \propto (\phi_J - \phi)^{-1}$,
is proportional to $(\phi_J-\phi)^{-3}$, irrespective of the spatial dimension.
The validity of this prediction has been confirmed by extensive
molecular dynamics simulations with soft disks.

However, one can note that this prediction differs from
other results on the divergence of the transport coefficients 
\cite{Garcia,Otsuki:PTP,Otsuki:PRE,Losert,Olsson}.
In particular, Garcia-Rojo et al. \cite{Garcia} concluded that the viscosity 
for two-dimensional monodisperse rigid-disks is proportional to 
$(\phi_\eta - \phi)^{-1}$,
where $\phi_\eta$ is the area fraction of the 2D order-disorder transition 
point, while the pressure diverges at a much higher $\phi_P$ with 
$p \propto (\phi_P-\phi)^{-1}$
\cite{Luding01,Luding01v2,Luding02,Luding04,Luding09}.  
Not only is the location of the divergence different, but also the power law
differs from the mean field prediction in Refs.\ \citen{Otsuki:PTP,Otsuki:PRE}.
{\em How can we understand these different predictions?}
One of the key points is that the situations considered are different from 
each other. As stated above, Garcia-Rojo et al. \cite{Garcia,Luding09} used 
two-dimensional monodisperse rigid-disks without or with very weak 
dissipation, whereas Otsuki and Hayakawa \cite{Otsuki:PTP,Otsuki:PRE} 
discussed sheared polydisperse granular particles with a soft-core potential
and rather strong dissipation.

In order to obtain an unified description on the critical behavior of the 
viscosity and the pressure in granular rheology, we numerically investigate 
sheared and weakly 
inelastic soft disks for both the monodisperse and the polydisperse 
particle size-distributions.  The organization of this paper is as follows:
In the next section, we summarize the previous estimates 
for the pressure and the viscosity for dense two-dimensional disk systems.
In Sec.\ \ref{Numerical:Sec}, we present our numerical results
for soft inelastic disks under shear in three subsections:
In Sec.\ \ref{Model:Subsec}, the numerical model is introduced, 
Sec.\ \ref{Mono:Subsec} is devoted to results on monodisperse systems,
and Sec.\ \ref{Poly:Subsec} to polydisperse systems.
In Sec.\ \ref{Time:Subsec}, a criterion for the ranges of validity of
the different predictions
about the divergence of the viscosity and the pressure is discussed.
We will summarize our results and conclude in Sec.\ \ref{Conclusion:Sec}.

\section{Pressure and viscosity overview}
\label{Previous:Sec}

In this section, we briefly summarize previous results on the behavior
of pressure and viscosity in two-dimensional disks systems.
Following Ref.~\citen{Luding09}, we introduce the non-dimensional 
%collisional 
pressure
\begin{eqnarray}
P^* & \equiv & P /(nT) - 1 ,
\label{P:def}
\end{eqnarray}
%XXX removed the 2 in E_kin/2N
where $P$ is the pressure, $n$ is the number density, 
and $T=\langle m (\bv{v} - \langle \bv{v} \rangle)^2 \rangle /(2N)$ is the 
%YYY added m - as is used in the rest of the paper formulation (p=nT...)
kinetic temperature (twice the fluctuation kinetic energy per particle per degree of freedom)
which is proportional to the square of the velocity fluctuations of each particle.
We also introduce the non-dimensional viscosity 
\begin{eqnarray}
\eta^* & = & \eta/(\rho v_T s_0/2)
\label{eta:def}
\end{eqnarray}
where $\rho$ denotes the particles' material density,
%YYY changed 'inner' to 'material' since I am not sure what inner means
$\rho^B=\rho\phi$ is the bulk area density, the fluctuation velocity
is denoted by $v_T = \sqrt{2T/m}$, $s_0=\sqrt{2 \pi}\sigma / 8$,
the mass of a grain (we assume all grains to have the same mass)
is denoted by $m$, and the mean diameter of a grain (disk) is denoted
by $\sigma$.
%with the particle material density $\rho=\rho^B/\phi$, with bulk area 
%density $\rho^B$, the thermal fluctuation velocity
%$v_T = \sqrt{2T/m}$, $s_0=\sqrt{2 \pi}\sigma / 8$, 
%the mass of each grain $m$ 
%and the averaged diameter $\sigma$ of particles. 
It should be noted that $\rho v_T s_0/2$ is the viscosity 
for a monodisperse rigid-disk system in the low-density limit
and correct to leading order in the Sonine polynomial expansion.
For later use, we also introduce the mean free time $t_E$ which is defined as 
the time interval between successive collisions.
This leads to the collision rate $t_E^{-1} = v_T \phi g(\phi) / s_0 = v_T/\lambda$ 
in the case of dilute and moderately dense systems of rigid disks,
where $\lambda$ is proportional to the mean free path.
%XXX please remove the lambda if you dont like it.

In the first part of this section, let us summarize previous results for 
elastically interacting rigid disk systems.
In the second part of this section, we show other previous results 
for soft granular disk systems under shear. 

\subsection{Rigid disk system in the elastic limit}
\label{Hard:Subsec}

For the equilibrium monodisperse rigid-disk systems,
%YYY added 'elastic' since the pre-factor 2 is only true for elastic
the reduced pressure $P^*$ of elastic systems at moderate densities 
$\phi < 0.67$ is well described by the classical Enskog theory
\cite{Fingerle,Luding01,Luding01v2,Luding04,Luding09}
\begin{equation}
P^*_4 = 2 \phi g_4(\phi). \label{P4:eq}
\end{equation}
with the aid of improved pair-correlation function at contact
\begin{equation}
g_4(\phi) = g_2(\phi) - \frac{\phi^3/16}{8(1-\phi)^4} ~,
\label{eq:g4}
\end{equation}
where $g_2(\phi) = \frac{1-7\phi/16}{(1-\phi)^2}$ in Eq.\ (\ref{eq:g4}) 
was proposed by 
Henderson in 1975 \cite{Henderson}.  
In the regime of high density $\phi > 0.65$, 
the reduced pressure becomes, first, lower than \eqref{P4:eq} because of 
ordering (crystallization) and, second, diverges at a density $\phi_P$ due
to excluded volume effects. This behavior is quantitatively fitted by 
\begin{equation}
P^*_{\rm dense} = \frac{2 \phi_P }{\phi_P - \phi} h(\phi_P - \phi) -1,
\end{equation}
with $\phi_P = \pi / (2 \sqrt{3})$, 
$h(x) = 1 + c_1 x + c_3 x^3$, and the fitting parameters
$c_1 = -0.04$, and $c_3 = 3.25$
\cite{Luding09,Luding01,Luding01v2,Herbst}.
As shown in references~\citen{Luding09,Luding01,Luding01v2} 
an interpolation law between the 
predictions for the low and the high density regions:
\begin{equation}
P^*_Q = P^*_4+ M(\phi) [P^*_{\rm dense} - P^*_4],
\label{P_Q}
\end{equation}
with $M(\phi) = [1 + \exp ( -(\phi - \phi_c)/m_0]^{-1}$,
$\phi_c = 0.699$, and $m_0 = 0.0111$, fits well
the numerical data for $P^*$. The quality of the empirical pressure 
function $P^*_Q$ is perfect, except for the transition region, for which 
deviations of order of 1\% are observed in the monodisperse, 
elastically interacting rigid disk system.

The dimensionless viscosity for monodisperse elastically colliding 
rigid disks 
is well described by the Enskog-Boltzmann equation
\begin{equation}
\eta^*_E = 
\left [ \frac{1}{g_2(\phi)}+ 2 \phi + 
\left (1 + \frac{8}{\pi} \right ) \phi^2 g_2(\phi) \right ].
\label{etaE:eq}
\end{equation}
%where Boltzmann-Enskog theory has been used for the derivation
%and $g_2(\phi)$ is the classical pair-correlation function at 
%contact, see Ref.\ \citen{Luding09} and the classical references therein.
Note that $g_2(\phi)$ satisfies
 $g_2(\phi) \approx g_4(\phi) \approx g_Q(\phi) 
 = P^*_Q / (2 \phi)$, for $ \phi \ll \phi_\eta$.
 A dominant correction, see Eq.\ (\ref{eta:eq}) below, controls the viscosity for 
higher densities, closer to $\phi \approx \phi_\eta$.

Equation \eqref{etaE:eq} can be used for low and moderate densities,
but it is not appropriate close to the crystallization area fraction $\phi_c$
\cite{Ishiwata,Garcia,Khain07,Khain09,Luding09,Otsuki:PTP,Otsuki:PRE}.
Therefore, an empirical formula for $\eta^*$ has been proposed as
\begin{equation}
\eta^*_L = \left 
(1 + \frac{c_\eta}{\phi_\eta - \phi} - \frac{c_\eta}{\phi_\eta} \right )\eta^*_E,
\label{eta:eq}
\end{equation}
which can fit the numerical data for $0 < \phi < \phi_\eta$
with two fitting parameters $c_\eta = 0.037$ and $\phi_\eta = 0.71$
\cite{Luding09}. Note that the last term is an improvement of the original
empirical fit \cite{Garcia} that makes $\eta^*_L$ approach unity for 
$\phi \rightarrow 0$. Note that  $\eta^*$
in Ref.\ \citen{Garcia} 
was obtained from a non-sheared system
by using Einstein-Helfand relation \cite{Helfand}.

A slightly different empirical form for the non-dimensional viscosity was
proposed by Khain \cite{Khain09} (based on simulations of a sheared system):
\begin{equation}
\eta^*_K = \left 
                (1 + \frac{c_\eta}{\phi_\eta - \phi} 
                     \left ( \frac{\phi}{\phi_\eta} \right )^3 \right ) 
                \eta^*_E ~,
\label{etaK:eq}
\end{equation}
with the same $c_\eta$ and $\phi_\eta$ as before.
The reasons for the difference between the viscosity in a sheared and 
a non-sheared system is an open issue and will not be discussed here.

We also introduce the scaled temperature given by
\begin{equation}
T^* = \frac{T(1-e^2)}{m \dot \gamma^2 s_0^2}
\end{equation}
for sheared inelastic rigid-disks, where $e$ and $\dot\gamma$ are 
the coefficient of restitution and shear rate, respectively.
Luding observed that the empirical expression 
\begin{equation}
T_K^* = \frac{\eta^*_K}{\phi^2 g_2(\phi)}
%\left [ \frac{1}{G(\phi)^2}+ \frac{2}{G(\phi)} +
%\left (1 + \frac{8}{\pi} \right )  \right ]
%\left ( 1 + \frac{c_\eta (\phi/\phi_\eta)^3}{\phi_\eta - \phi}
%- \frac{c_\eta}{\phi_\eta} \right )
\label{TK}
\end{equation}
fits best the numerical data for monodisperse rigid disks \cite{Luding09},
while 
\begin{equation}
T_L^* = \frac{\eta^*_L}{\phi^2 g_2(\phi)}
%\left [ \frac{1}{G(\phi)^2}+ \frac{2}{G(\phi)} +
%\left (1 + \frac{8}{\pi} \right )  \right ]
%\left ( 1 + \frac{c_\eta (\phi/\phi_\eta)^3}{\phi_\eta - \phi}
%- \frac{c_\eta}{\phi_\eta} \right )
\label{TL}
\end{equation}
slightly overpredicts the scaled temperature.

For polydisperse elastic rigid-disk systems,
many empirical expressions for the reduced pressure $P^*$
have been proposed, 
see e.g.\ \cite{Luding09,Luding01,Luding02,Luding04,Torquato95}.
It is known that $P^*$ diverges around $\phi_{\rm max} \simeq 0.85$ for bi- and 
polydisperse systems, but there is no theory to our knowledge that predicts 
the dependence of $\phi_{\rm max}$ on the width of the size distribution 
function that was observed in rigid-disk simulations \cite{Luding01,Luding02}.
Dependent on the dynamics (rate of compression), on the material parameters
(dissipation and friction), and on the size-distribution, different
values of $\phi_{\rm max}$ can be observed.
In several studies, the critical behavior was well described
asymptotically by a power law 
\begin{equation}
P_d^* \sim (\phi_{\rm max} - \phi)^{-1}
\label{Pd:eq}
\end{equation}
see Refs.\ \citen{Torquato95,Luding01,Luding02}.
%which can also be obtained from a free volume theory for the 
%ordered phase of the monodisperse systems \cite{Luding01},
%where $\phi_{\rm max}$ is treated as a fitting parameter.,

No good empirical equation for the viscosity of polydisperse rigid-disk 
systems in the elastic limit has been proposed to our knowledge.
However, if we assume that the viscosity behaves like that of
the monodisperse rigid-disk system, we can introduce the empirical 
expression 
\begin{equation}
\eta_{d}^* \sim (\phi_{\rm max} - \phi)^{-1}
\label{etad:eq}
\end{equation}
as a guess. 
Here, we assume that 
the pressure $P^*$ and the viscosity $\eta^*$ for the polydisperse system
diverge at the same point $\phi_{\rm max}$,
which differs from the case of the monodisperse system,
where $P^*$ and $\eta^*$ diverge at different points 
$\phi_P$ and $\phi_\eta$ due to the ordering effect.

\subsection{Soft-disk system}
\label{Soft:Subsec}

Let us consider a sheared system of inelastic soft-disks characterized by
 the non-linear normal repulsive contact force $k\delta^\Delta$ with 
 power $\Delta$, where $k$ and $\delta$ are the stiffness constant and 
the compression length (overlap), respectively. 
For this case, Otsuki and Hayakawa~\cite{Otsuki:PTP,Otsuki:PRE} 
proposed scaling relations for the kinetic temperature $T$, 
shear stress $S$, and pressure $P$, near the jamming 
transition point $\phi_J \simeq 0.85$:
\begin{equation} \label{scaling}
T  =  |\Phi|^{x_{\Phi}} {\cal T}_{\pm}\left(\frac{\dot\gamma}{|\Phi|^{\alpha}}\right),  \
S  =  |\Phi|^{y_{\Phi}}{\cal S}_{\pm}\left(\frac{\dot\gamma}{|\Phi|^{\alpha}}\right),  \
P  =  |\Phi|^{y_{\Phi}'}{\cal P}_{\pm}\left(\frac{\dot\gamma}{|\Phi|^{\alpha}}\right), 
\end{equation}
where $\Phi \equiv \phi - \phi_J$ is the density difference from
the jamming point. 
This scaling ansatz is based on the idea that the system has only one
relevant time-scale $\tau \sim |\Phi|^{-\alpha}$
diverging near the transition point $\phi_J$,
and the behavior of the system is dominated by the ratio
of the time scale $\tau$ and the inverse of the shear rate $\dot \gamma$.
This idea is often used in the analysis of critical phenomena.

The scaling functions
${\cal T}_{+}(x)$, ${\cal S}_{+}(x)$, and ${\cal P}_{+}(x)$
satisfy 
\begin{eqnarray}\label{jammed}
\lim_{x\rightarrow 0}{\cal T}_+(x) & =& x, \quad 
\lim_{x\rightarrow 0}{\cal S}_{+}(x) = 1, \quad 
\lim_{x\rightarrow 0}{\cal P}_{+}(x) = 1
\end{eqnarray}
for $\phi > \phi_J$, i.e., for higher area fraction.
% since {\bf ... say in a half-sentence why? ...}
The pressure and shear stress scaling -- in this limit -- 
represent the existence of a (constant) yield stress $S=S_Y$.  
The scaling for the temperature is obtained from the assumption
that a characteristic frequency, $\omega \equiv \dot \gamma S / (nT)$,
%which is proportional to the characteristic frequency for the density of state
%in the jammed state $\phi > \phi_J$ \cite{Wyart},
is finite when $\dot \gamma \to 0$ in the jammed state 
$\phi > \phi_J$, see Ref.\ \citen{Wyart}.
\footnote{Here, we should note that $\omega$ is proportional to 
the Enskog collision rate $\omega = (1-e^2) t_E^{-1} /2$,
see Ref.\ \citen{Luding09}, in the unjammed state well below the
jamming point, $\phi <\phi_J$, i.e., in the collisional flow regime. 
%XXX
Due to the prefactor $(1-e^2)/2$, we can identify 
$\omega$ with the characteristic dissipation rate.
%{\bf Different time-scales (inverse frequencies) and their relative
%importance are discussed below in subsection \ref{subsubsec:dimensionless}.}
The different time-scales (inverse frequencies) and their relative
importance are discussed below in subsection \ref{subsubsec:dimensionless}.
}

On the other hand, for lower area fraction, 
${\cal T}_{-}(x)$, ${\cal S}_{-}(x)$, and ${\cal P}_{-}(x)$
satisfy
\begin{eqnarray}\label{unjammed}
\lim_{x\rightarrow 0}{\cal T}_{-}(x) & = & x^2, {\quad}
\lim_{x\rightarrow 0}{\cal S}_{-}(x) = x^2,  \quad 
\lim_{x\rightarrow 0}{\cal P}_{-}(x)  =  x^2
\end{eqnarray}
for $\phi \ll \phi_J$, 
which represent Bagnold's scaling law in the liquid phase.

Furthermore, for diverging argument $x$, i.e., at the jamming
point J with $\Phi \rightarrow 0$, the scaling functions
${\cal T}_{\pm}(x)$, ${\cal S}_{\pm}(x)$, and ${\cal P}_{\pm}(x)$
should be independent of $\Phi$ and thus satisfy:
\begin{equation}\label{critical}
\lim_{x\rightarrow \infty}{\cal T}_{\pm}(x) = x^{x_\Phi/\alpha}, \
\lim_{x\rightarrow \infty}{\cal S}_{\pm}(x) = x^{y_\Phi/\alpha}, \
\lim_{x\rightarrow \infty}{\cal P}_{\pm}(x) = x^{y_\Phi'/\alpha} \,.
%\lim_{x\rightarrow \infty}{\cal W}_{\pm}(x) = x^{z_\gamma}
\end{equation}
%{\bf I removed the $\alpha$-prime' in (2.18) and I assume we remove this 
%sentence fragment: "which represents the definition of the exponents with
%subscripts $\gamma$."}

The critical exponents in Eq.(\ref{scaling}) are given by
\begin{equation}
\label{exponents}
x_{\Phi} = 2+\Delta, \quad 
y_{\Phi} = y_{\Phi}' = \Delta, \quad 
{\rm and~} \quad \alpha=\frac{\Delta+4}{2},
\end{equation} 
which depend on some additional assumptions\cite{Otsuki:PTP},
such as the requirement that the pressure $P$ for $\dot \gamma \to 0$, 
in the jammed state $\Phi >0$, scales with the force power-law as 
$P \sim \Phi^{\Delta}$, see Refs.\ \citen{OHern,OHern03}.

Thus, the temperature $T$, the shear stress $S$, and the pressure $P$, 
below the jamming transition point in the zero shear limit 
$\dot \gamma \to 0$ obey:
\begin{equation} 
\label{scale_T_S_P}
T \sim (\phi_J - \phi)^{-2} \dot \gamma ^2,
\quad
S \sim (\phi_J - \phi)^{-4} \dot \gamma ^2,
\quad
P \sim (\phi_J - \phi)^{-4} \dot \gamma ^2.
\end{equation}
Both the viscosity $\eta=S/\dot\gamma$ and pressure $P$, at the jamming 
transition point, diverge proportional to the area fraction difference 
to the power $-4$.
Substituting Eqs.\ \eqref{scale_T_S_P} into 
Eqs.\ \eqref{P:def} and \eqref{eta:def}, the reduced pressure $P^*$
and the dimensionless viscosity $\eta^*$, in the vicinity of the jamming 
point are respectively given by
\begin{eqnarray}
P_{J}^*  & \sim &  (\phi_J - \phi)^{-2},  \label{scaling:P}\\
\eta_{J}^* & \sim &  (\phi_J - \phi)^{-3}. \label{scaling:eta}
\end{eqnarray}

It is remarkable that the scaling relations 
\eqref{scale_T_S_P}--\eqref{scaling:eta} 
below the jamming transition point
are independent of $\Delta$, even
though the exponents in Eq.\ \eqref{exponents}
depend on $\Delta$.
The validity of Eqs.\ \eqref{scaling:P} and \eqref{scaling:eta} 
for various $\Delta$ has been numerically verified \cite{Otsuki:PTP,Otsuki:PRE}.
However, the conjecture that the scaling relations
\eqref{scaling:P} and \eqref{scaling:eta} are applicable
in the hard disk limit seems to be  in conflict with the empirical relations 
Eqs.\ \eqref{Pd:eq} and \eqref{etad:eq} for elastic rigid-disk systems. 

\section{Numerical results}
\label{Numerical:Sec}

In this section, we numerically investigate the reduced
pressure $P^*$ and viscosity $\eta^*$ of sheared systems with
soft granular particles, with special focus on the rigid-disk limit.
In the first part, our soft-disk model is introduced. 
In the second part, we present numerical results for monodisperse systems,
while in the third part the results for polydisperse systems are presented.

\subsection{The soft-disk model system}
\label{Model:Subsec}

\subsubsection{Contact forces and boundary conditions}

Let us consider two-dimensional granular assemblies under a uniform shear
with shear rate $\dot \gamma$.
%The packing fraction of the system is denoted by  $\phi$.
Throughout this paper, we assume that granular particles are frictionless, 
without any tangential contact force acting between grains. For the sake
of simplicity, we restrict ourselves to the linear contact model
with $\Delta=1$.
We assume that all particles have identical mass regardless of their diameters. 
The linear elastic repulsive normal force between the grains $i$ and $j$, 
located at $\bv{r}_i$ and $\bv{r}_j$, is:
%The elastic force is given by
\begin{eqnarray}
f_{\rm el}(r_{ij}) & = & k \Theta 
                             \left( \sigma_{ij} - r_{ij} \right )
                         (\sigma_{ij}-r_{ij}),
\label{elastic:force}
\end{eqnarray}
where $k$ and $r_{ij}$ are the elastic constant and the distance between 
the grains $r_{ij}\equiv |\bv{r}_{ij}|=|\bv{r}_i-\bv{r}_j|$, respectively.
$\sigma_{ij} = (\sigma_i + \sigma_j)/2$ is the average of the diameters
of grains $i$ and $j$.
The Heaviside step function $\Theta(x)$ satisfies $\Theta(x) = 1$
for $x \geq 0$ and $\Theta(x) = 0$ otherwise.
The viscous contact normal  force is assumed as
\begin{eqnarray}
f_{\rm vis}(r_{ij}, v_{ij,{\rm n}}) & = & 
- \zeta \Theta \left( \sigma_{ij} - r_{ij} \right ) 
v_{ij,{\rm n}},
\label{dis:lin}
\end{eqnarray}
where $\zeta$ is the viscous parameter. Here, $v_{ij,{\rm n}}$ is
the relative normal velocity between the contacting grains 
$v_{ij,{\rm n}}\equiv (\bv{v}_i-\bv{v}_j)\cdot\bv{r}_{ij}/r_{ij}$,
where $\bv{v}_i$ and $\bv{v}_j$ are the velocities of the centers of the 
grains $i$
and $j$, respectively.

In order to obtain a uniform  
velocity gradient $\dot\gamma$ in  
$y$ direction and macroscopic velocity only 
in $x$ direction, we adopt
the Lees-Edwards boundary conditions. 
The average velocity $\bv{c}(\bv{r})$ at position $\bv{r}$ 
is given by $\bv{c}(\bv{r}) = \dot \gamma y \bv{e}_x$, where 
$e_{x,\alpha}$ is a unit vector component given by
$e_{x,\alpha} = \delta_{x\,\alpha}$, where $\alpha$ is the Cartesiani coordinate.
%YYY changed the wording above

\subsubsection{Discussion of dimensionless quantities}
\label{subsubsec:dimensionless}

There are several non-dimensional parameters in our system.
One is the restitution coefficient $e$ given by
\begin{equation}
e \equiv \exp \left [-\frac{\pi \zeta}{\sqrt{2 k/m - (\zeta/m)^2}} \right ] 
       = \exp \left [ - \zeta t_c \right ] ~,
\end{equation}
with the pair-collision
\footnote{The contact duration $t_c$ is well defined for two masses connected by
a linear spring-dashpot system and corresponds to their half-period of oscillation.
A particle in a dense packing (connected to several masses by linear spring-dashpots)
has a somewhat higher oscillation frequency, but the order of magnitude 
remains the same. Particles with non-linear contact models can have 
a pressure dependent $t_c$, but are not considered here.
%YYY re-phrased this footnote again ...
%XXX Here would be a good place to discuss this - but on the other hand
%XXX we don't show any nonlinear forces results ... so we can leave it out.
%If we use the non-linear elastic force $k \delta^\Delta$ 
%and the viscous force $\zeta \delta^{\Delta-1} v_n$
%with the compression length $\delta$
%and the relative velocity $v_n$,
%we cannot estimate the dimensionless
%contact duration $\tau_c^*$ and the restitution
%coefficient $e$ because the contact duration
%$t_c$ and $e$ depends on the relative velocity
%at the collision.
%However, even in the case of the non-linear forces,
%we expect  similar results to the case of the linear forces
%by introducing the dimensionless time for the elastic forces
%$\tau_k^* = \tau_k \dot \gamma$
%and the dimensionless viscous constant
%$\zeta^* = \zeta / (m \tau_k^{-1} \sigma^{1-\Delta})$
%with the characteristic time $\tau_k \equiv \sqrt{m \sigma^{1-\Delta} / k}$
%for the elastic force,
%which correspond to $\tau_c^*$ and $1-e^2$, respectively.
} 
contact duration $t_c \equiv \pi / \sqrt{2 k/m - (\zeta/m)^2}$.
%which is inverse proportional to the stiffness. 
Another is the dimensionless contact duration  
\begin{equation}
\tau_c^* \equiv t_c \dot \gamma
\label{tau:def}
\end{equation}
that represents the ratio of the two ``external'' time-scales of the system
\footnote{One can see $\tau^*_c = (\sigma \dot \gamma) / (\sigma /t_c)$ also 
as the ratio of the two relevant velocities in the dense limit, i.e.,
as the ratio of the local velocity of horizontal layers that are a 
diameter of a grain, $\sigma$, apart,
and the local information propagation speed $\sigma/t_c$ in a dense packing.
%{\bf With this I am now really convinced that this is the best
%dimensionless material parameter - even though it is always small.}
%{\bf But the ratio of velocities makes only sense in the dense regime,
%since $t_c$ is NOT the relevant time-scale in the dilute+hard regime!}
However, the ratio of velocities makes only sense
in the dense, soft regime, since $t_c$ is not a relevant time-scale
in the dilute, near-rigid regime.
}.
``External'' means here that these time scales are externally 
controllable, i.e., the contact-duration is a material parameter and
the inverse shear rate is externally adjustable.

In all cases studied later, we have $\tau^*_c \ll 1$, which means
that the shear time scale is typically much larger than the contact
duration, i.e., we do {\em not} consider the case of very soft particles,
which is equivalent to extremely high shear rates.  
Therefore, $\tau^*_c$ will be used as dimensionless
control parameter in order to specify the magnitude of stiffness:
%The smaller (larger) $\tau^*_c$ the harder (softer) the particles are 
The rigid disks are reached in the limit $\tau^*_c \to 0$.
%-- with the hard-disk limit case reached for very near-rigid disks $\tau^*_c \to 0$.
%{\bf The above line was wrong in the previous version -- I hope I changed
%it to be correct. Please check.}
%\footnote{Another, related dimensionless quantity is
%$\tau^*_L = (L/d) t_c \dot \gamma = t_c (L/d) (v_s/L) = t_c v_s / d$,
%which represents the ratio of the time it takes information to propagate 
%in a dense system over a distance of the system size $L$, and 
%the shear-time $v_s/L$. {\bf drop this footnote - only for discussion and 
%maybe later use ....}}

The third time-scale, $t_E$, in the system is an ``internal'' variable,
i.e., cannot be controlled directly. This time scale is proportional to 
the inverse characteristic frequency of interactions, %$\omega^{-1}$, with $\omega^{-1} \sim $
i.e., the mean free time, $t_E$, in the dilute case or the rigid-disk 
limit. This defines the (second) dimensionless ratio of times
\begin{equation}
\tau_E^*  \equiv t_E \dot \gamma 
%          = \frac{1-e^2}{2} \, \frac{\dot \gamma}{\omega}
%          = \frac{(1-e^2)nT}{2S} ~,
\label{tauE:def}
\end{equation}
relevant in the dilute, collisional regime.

The third dimensionless number is defined as the ratio of
contact duration and mean free time,
\begin{equation}
\tau_{cE}^* \equiv \frac{t_c}{t_E} 
%       = \frac{2t_c \omega}{(1-e^2)} 
       = \frac{\tau^*_c}{\tau^*_E} ~.
\label{tauc:def}
\end{equation}
see Eq.\ (53) in Ref.\ \citen{Luding09}.
%{\bf Note that I changed all $\tau_{CE}$ to $\tau_{cE}$ for consistency
%of the subscripts ... maybe one or the other figure has to be changed
%accordingly? Michio, can you please take care -- if you both agree with
%this nomenclature?}
The meaning of this dimensionless number is as follows: For 
very small $\tau_{cE}^* \ll 1$ one is in the binary collision regime, for 
large $\tau_{cE}^* > 1$, one is in the solid-like regime with long-lasting
multi-particle contacts. 
In the hard disk limit $\tau_c^*\to 0$, 
we can identify $\tau_{cE}^*$ with the coordination number as will be shown in Fig.\ \ref{fig:Z}. 
Namely, finite $\tau_{cE}^*$ in the near-rigid situation means that the system is in a jammed phase.
% changed above - correct?

The binary collision regime, $\tau_{cE}^* \rightarrow 0$,
cannot be controlled directly, since $t_E$ is a function
of temperature, which depends on $e$ and $\dot \gamma$.
On the other hand,
the rigid-disk limit, $\tau_{c}^* \to 0$, %and $e=const.$,
can be approached/realized by either
($i$) vanishing shear rate, $\dot \gamma \to 0$, or
($ii$) near-rigid particles with high stiffness, $k \to \infty$
(with controlling the variable $\zeta$ to maintain a constant restitution coefficient $e$).

%{\bf 
%Michio: I have changed $\tau_c$ to $\tau_{cE}^*$, where I add $*$
%in order to show that $\tau_{cE}^*$ is a non-dimensional quantity.
%}
 
\begin{table}[htb]
\begin{tabular}{|l|l|l|l|c|c|c|c|}
\hline
                  & ratio of times            &  ratio of velocities %XXX
                                                          / stresses  
                                               & regime of relevance\\
\hline
\hline
$\tau_c^*$        & $t_c / \dot \gamma^{-1}$  & $v_{\sigma\dot\gamma}/v_c 
                                                = \sigma \dot \gamma / (\sigma /t_c)$
                  & near-rigid, high density ($\sigma \gg \lambda$, $t_c \gg t_E$)\\
\hline
\hline
$\tau_E^*$        & $t_E / \dot \gamma^{-1}$  
                  & $v_{\lambda\dot\gamma}/v_E = \lambda \dot \gamma / (\lambda/t_E)$ 
                  & rigid, low density ($\sigma \ll \lambda$, $t_c \ll t_E$)\\
$\tau_{cE}^*$     & $t_c / t_E $              & $v_E/v_c = \sigma/t_E / (\sigma/t_c)$
                  & near-rigid, low and moderate densities  \\
%\hline
%$\tau_{s}^*$      & $[t_E + t_c] / \dot\gamma^{-1}$  & $v_{(\sigma+\lambda)\dot\gamma}/v_s$ 
%                                                       (see main text)
%                  & hard and rigid, all densities  \\
\hline
%\hline
%old:  & {\bf REMOVE} \\ %XXX
%\hline
%$\tau_\omega^*$        & $\omega^{-1} / \dot \gamma^{-1}$  
%                  & $v_{\sigma\dot\gamma}/v_{\omega}=\sigma\dot\gamma/(\sigma \omega)$
%                  & rigid, low density ($\sigma \ll \lambda$, $t_c \ll t_E$)\\
%$\tau_{c\omega}^*$     & $t_c / \omega^{-1} $  
%                  & $v_\omega/v_c=\sigma\omega/(\sigma/t_c)$
%                  & hard, all densities  \\
%\hline
%$\tau_{s\omega}^*$     & $[\omega^{-1}+t_c] / \dot\gamma^{-1}$  & $~$ (see main text)
%                  & all densities  \\
%\hline
%\hline
%new?:\\
\hline
$\tau_\omega^*$        & $\omega^{-1} / \dot \gamma^{-1}$  
%XXX old          & $v_{\sigma\dot\gamma}/v_{\omega}=\sigma\dot\gamma/(\sigma \omega)$
                  & $nT/S = 2\tau_E^* / (1-e^2)$
                  & well defined in sheared systems\\
%XXX in the new version, I think we dont need $\tau_{c\omega}^*$???
$\tau_{c\omega}^*$     & $t_c / \omega^{-1} $  
%XXX old          & $v_\omega/v_c=\sigma\omega/(\sigma/t_c)$
                  & $t_c \dot\gamma S/(nT) = \tau_c^* / \tau_\omega^*$
                  & well defined in all systems\\
%\hline
%$\tau_{s\omega}^*$     & $[\omega^{-1}+t_c] / \dot\gamma^{-1}$  & $~$ (see main text)
%                  & \\
\hline
\end{tabular}
\caption{
Summary of the dimensionless numbers discussed in the text, where 
$t_c$, $\dot \gamma^{-1}$, $t_E$, $\omega^{-1}$ are
contact duration, inverse shear rate, mean free time,
and inverse characteristic %XXX frequency,
dissipation rate,
respectively.  The velocities 
$v_{L\dot\gamma}$, $v_c$, and $v_E$ %, $v_\omega$, and $v_s$ 
are
the shear velocity of layers separated by length $L$, 
the speed of sound propagation in a dense packing, and
the speed of sound propagation in a dilute packing, 
%the speed related to the dissipation rate, %XXX this is not a good wording,
% but I could not identify what this means in terms of speed -- I do not see
% that $v_\omega$ does correspond to a speed at all.
% Maybe we shold better use the stress-ratios when talking about the
% \tau_\omega?
% XXX - if we chose for the stress-ratios in the table, $v_\omega$ can be removed
%and the general sound propagation speed, 
respectively.
The relevant lengths $L$ can be the diameter $\sigma$ (in the dense limit), 
the mean free path $\lambda=\lambda(\phi)$ (in the dilute limit), 
or their sum (for all densities). 
%I DO NOT TOUCH THE DEFINITION OF $\lambda$ HERE. IF YOU HAVE A BETTER DEFINITION, PLEASE REPLACE IT.
}
\label{tab:tab1}
\end{table}

Furthermore, we can introduce dimensionless numbers that are
related to the inverse characteristic dissipation rate
$\omega^{-1}$ 
\footnote{Note that the identity 
$\omega^{-1} = 2t_E/(1-e^2)$ is true in the dilute, collisional limit
only. For higher densities and for softer particles, one has 
$\omega^{-1} > 2 t_E/(1-e^2)$, i.e., energy dissipation becomes 
somewhat slower when approaching the jamming transition.
This is consistent with a slower energy decay 
%reported as ``detachment'' effect in Ref.\ \cite{luding94d} and with 
due to the reduced dissipation rate, proposed in Eq.\ (52) 
in Ref.\ \citen{Luding09}},
which has the meaning of the energy dissipation time-scale.
For $e \rightarrow 1$, dissipation is becoming very slow, while
for small $e \sim 0$, considerable energy can be dissipated, 
within a time of order of $t_E$ or $t_c$.

Replacing $t_E$ by $\omega^{-1}$ in Eqs.\ (\ref{tauE:def}) and
(\ref{tauc:def}), we obtain
\begin{eqnarray}
\tau_\omega^* & \equiv & \dot \gamma/\omega \,,
\label{eq:tauw} \\
\tau_{c\omega}^{*} & \equiv & t_c \omega \,.
\label{taucEd:def}
\end{eqnarray}
It should be noted that
$\tau_\omega^*$ and $\tau_{c\omega}^*$ approximately satisfy 
the relations $\tau_\omega^* \approx 2\tau_E/(1-e^2)$ %XXX Check - this I changed, correct?
and $\tau_{c\omega}^{*} \approx (1-e^2)\tau_{cE}^*/2$, respectively,
in the collisional regime,
where the prefactor plays an important role,
as will be demonstrated later. 
 
%{\bf === REMOVE FROM HERE ===\\
%after I had written this text, I had again a boost in understanding,
%just remove the following paragraph - I think no longer it makes sense.
%$\tau_s^*$ does make sense above, however, so please keep it.}
%In analogy to $\tau_s^*$, that involves the interaction rate,
%we can introduce
%\begin{equation}
%\tau_{s\omega}^* 
%         = \frac{2 \tau_s^*}{ (1-e^2) }
%         = \left [ 1/\tau_{cE}^* + 1
%           \right ] \frac{2 \tau_c^*}{ (1-e^2) } ~,
%\label{eq:tausw}
%\end{equation}
%%TODO CHECK ...
%which is relevant for all densities, and combines the intuitive dimensionless number
%$\tau_{cE}^*$ that distinguishes between binary and multi-particle collisional regime,
%and the dimensionless stiffness parameter $\tau_c^*$ multiplied with the inverse 
%dissipation factor $2/(1-e^2)$.
%In Eq.\ (\ref{eq:tausw}), the dimensionless numbers 
%$2\tau_c^*/(1-e^2)$ and $\tau_\omega^*$
%are realized as the high and low density (soft and hard particle)
%limit of $\tau_{s\omega}^*$, respectively.
%{\bf === REMOVE UNTIL HERE ===}

The consequences of the interplay among these dimensionless numbers will be clarified 
and discussed in the following sections.  Furthermore, we will identify the dimensionless
number that -- we believe -- allows us to distinguish between the two scaling regimes.

\subsubsection{Simulation parameters}

We examine two systems with different grain diameters and composition.
The first {\it monodisperse} system consists of only one type of particles,
whose diameters are $\sigma_0$.  The other {\it polydisperse} system consists 
of two types of grains, and the diameters of grains are 
$0.5\sigma_0$, and $\sigma_0$, where 
the numbers of each type of grains are $0.8N$ and $0.2N$, respectively,
with the total number of particles $N$.
The reasons to study such a polydisperse system
are (i) to avoid crystallization 
and (ii) to compare our new near-rigid data with previous 
results from rigid disks \cite{Luding01,Luding02}.

In our simulations, the number of particles is $N=2401$ except for
the data in Figs.\ \ref{P_eta_poly_e0.9} and \ref{P_eta_poly_log_e0.9},
where we have used $N=20000$.
We use the leap-frog algorithm, which is second-order accurate in time, 
with the time interval $\Delta t=0.2\sqrt{m / k}$. We checked that the 
simulation converges well by comparison with a shorter time-step 
$\Delta t=0.02\sqrt{m / k}$.

The pressure and the viscosity are respectively given by 
%
%YYY in the equation for pressure below I first removed the '2'
% since the sum does not go over all pairs, but only j>i
% but then I realized that the 2 is needed because of the 
% definition of pressure ... thus nothing changed ...
% please confirm - true?
%
\begin{eqnarray}
P & = & 
\frac{1}{2V} \left < \sum_{i=1}^N \sum_{j>i} r_{ij}
 \left [ f_{{\rm el}}(r_{ij}) + 
f_{{\rm vis}}(r_{ij}, v_{ij,{\rm n}}) \right ]
+
 \sum_{i=1}^N \frac{|\bv{p}_i|^2}{m} \right >, 
\label{P:ex}\\
\eta & = &  -\frac{1}{\dot \gamma V}\left <    
        \sum_{i=1}^N \sum_{j>i} \frac{r_{ij,x} r_{ij,y}}{r_{ij}}
 \left [ f_{{\rm el}}(r_{ij}) + 
f_{{\rm vis}}(r_{ij}, v_{ij,{\rm n}}) \right ]
+   \sum_{i=1}^N \frac{p_{i,x}p_{i,y}}{m} \right >
\label{S:calc},
\end{eqnarray}
with the volume of the system $V$, the relative distance vector
${\bv{r}}_{ij}=(r_{ij,x},\,r_{ij,y})$, with $r_{ij}=|{\bv{r}}_{ij}|$,
 and the peculiar momentum
$\bv{p}_i = (p_{i,x},\,p_{i,y})
          \equiv m(\bv{v}_i - \dot\gamma y_i \bv{e}_x)$.

\subsection{Mono-disperse system}
\label{Mono:Subsec}

\begin{figure}
\begin{center}
\includegraphics[height=14em]{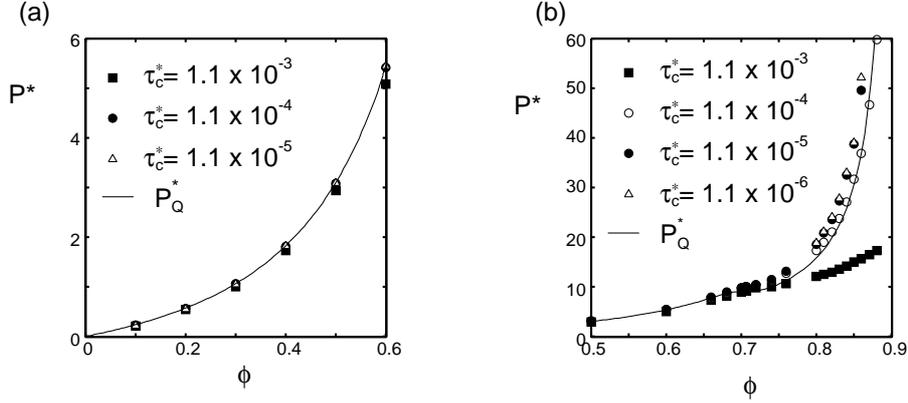}
\caption{ 
  The reduced pressure $P^*$ as a function of the area fraction
  $\phi$ in the {\it monodisperse} system with $e=0.999$, for different
  $\tau_c^*$, as given in the inset, 
  %$ = 1.11 \times 10^{-3}$, $1.11 \times 10^{-4}$, $1.11 \times 10^{-5}$, 
  %and $1.11 \times 10^{-6},$ 
  and for $\phi<0.6$ (a) and $\phi > 0.5$ (b).
 }
\label{P_hard}
\end{center}
\end{figure}

In Figs.\ \ref{P_hard}(a) and (b), 
we plot $P^*$ as a function of the area fraction
$\phi$ in the {\it monodisperse} system with $e=0.999$ 
for $0 < \phi < 0.6$ and $0.5 < \phi < 0.9$, respectively.
Most of all data of $P^*$ seem to converge in
the rigid-disk limit ($\tau_c^* \to 0$).
Moreover,  the data for $P^*$ with $\phi<0.6$ 
are consistent with $P^*_Q$, see Fig.\ \ref{P_hard}(a),
while $P^*$ for $\phi>0.7$ in Fig.\ \ref{P_hard}(b)
deviates from $P_Q^*$ in the soft case of $\tau_c^*=1.11 \times 10^{-3}$,
and also in the rigid-disk limit.  Only the simulations
with $\tau_c^*=1.11\times 10^{-4}$ are close to $P_Q^*$
-- seemingly by accident.  At high densities, 
for very soft particles, the stress is considerably smaller 
than predicted by $P^*_Q$, while for near-rigid particles, we 
observe a higher stress.

%{\bf Stefan : to be discussed further - maybe the zero-shear data 
%solve this discrepancy? Then we can write:
%"The pressure from simulations without shear indeed agrees with
%$P_Q^*$, so that we can make the presence of shear responsible
%for the higher stress in the hard-disk limit in Fig.\ \ref{P_hard}(b)."}
%
%{\bf Michio : As shown in the data I have sent, $P^*$ in the 
%monodisperse system
%for $e=0.999$ without shear (freely cooling state) deviates from
%$P^*_Q$. So, I conjecture that the origin of 
%the discrepancy is not the effect of the shear flow, but the inelasticity 
%of the particles. I think that the fact that $P^*$ for $e=0.999$
%coincides with $P^*_Q$ supports this conjecture.}

\begin{figure}
\begin{center}
\includegraphics[height=14em]{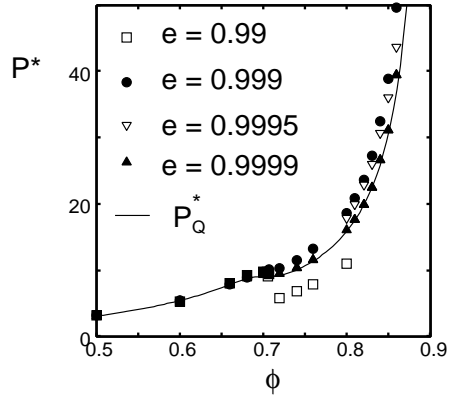}
\caption{
  %{\bf If you decide to leave the snapshot, please add (a) and (b),
  %and change the snap-figure, like Fig.3, otherwise remove it from the
  %caption.}
  %(a) The reduced pressure $P^*$ as a function of the area fraction
  The reduced pressure $P^*$ as a function of the area fraction
  $\phi$ in the {\it monodisperse} system with
  $\tau_c^* = 1.11 \times 10^{-5}$ and different $e$,
  as given in the inset. %$e=0.99$, $0.999$, $0.9999$.
  %(b) The snapshot is from the simulation with $e=0.999$
  %and $\dot \gamma = 0$. {\bf True? (from your last mail) maybe better
  %use a snapshot from the sheared case?}
 }
\label{P_elastic}
\end{center}
\end{figure}

In order to check the possibility that the restitution coefficient is
the reason for the deviation between the numerical data and $P_Q^*$ in 
Fig.\ \ref{P_hard}(b),
we plot $P^*$ for different $e$, for $\tau_c^*=1.11\times 10^{-5}$
in Fig.\ \ref{P_elastic}.
%The data for $e=0.99, 0.999$ deviate from $P_Q$, but 
%the data is on the curve for $P_Q^*$ in the almost elastic 
%case $e=0.9999$.
At high densities, 
for inelastically interacting particles, $e=0.99$, the stress is considerably smaller 
than predicted by $P^*_Q$, while for more elastic particles, we 
observe a higher stress. Only the almost elastic case $e=0.9999$
is close to the prediction.

%{\bf Stefan : I better understand now - thanks for all the explanations!  
%But I have a small
%problem with the irregular transition: e=0.99 data are below, then e=0.999 data
%are above and then e=0.9999 data are on PQ* ... actually, e=0.999, since there
%is no shear-band, should agree with PQ* - unless the shear leads to some other
%effects ... we will see that when we look at some non-sheared data.
%Anyway the question remains: when going from $e=0.9$ in small steps to
%$e=0.9999$, for one density and e.g. $k*=5\times 10^5$, is it a smooth
%transition? or is it sharp???}

The low pressure for $e=0.99$ is due to the existence of a shear-band 
-- see below. 
For all other situations, no shear-band is observed, 
%{\bf True?}
however, different patterns of defect lines in the crystal are evidenced
for $e=0.9990$ and $e=0.9995$, while an almost perfect crystal is observed
for $e=0.9999$, where slip-lines appear.
%{\bf Are the slip-lines (shear-bands of width $W=d$?) always at the same
%position or do they jump? or move? Please add 1-2 sentences about this.}
It should be noted that the positions of the 
slip-lines (shear-bands of width $W=d$) don't move 
in the steady state of one sample,
but vary among different samples.
%{\bf True? Please check that I understood correctly.
%Here a new question: how does shear happen in the almost elastic
%cases? Is there a slip-line (always at the same position)? or is
%the shear distributed through the system? homogeneously? or
%intermittent?}
%
%So, I think that a sharp transition is observed between $e=0.99$ and 
%$e=0.9999$. But, we can observe smooth transition if we go from $e=0.999$ 
%to $e=0.9999$ because there is no shear-band in this region.
%In order to verify this fact, I have added the data for $e=0.9995$
%in Fig.\ \ref{P_elastic}, which supports my conjecture.
%}

\begin{figure}
\begin{center}
\includegraphics[height=14em]{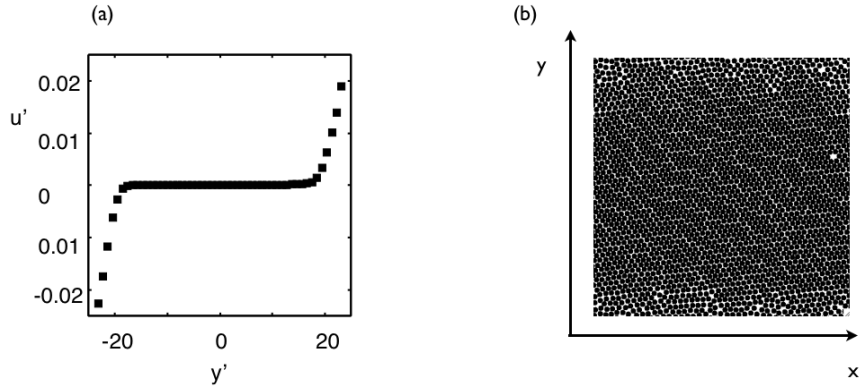}
\caption{ 
  (a) The scaled velocity $u'=u/(\sigma_0 / \sqrt{m/k})$ 
  in $x$ direction as a function of $y'=y/\sigma_0$,
  for $\phi=0.84$, $\tau_c^* = 1.11 \times 10^{-5}$, and 
  $e=0.99$. 
  (b) Snapshot of the {\it monodisperse} system from (a).
% for $\phi=0.76$, $\tau_c^* = 1.11 \times 10^{-5}$, and 
% $e=0.99$.
%{\bf The density 0.84 is not consistent with Figs.\ 4 and 5. can you 
%print a shear-band at density 0.76 (like in Fig.4 and 5) here?}
 }
\label{grad_snap_mono_e0.99}
\end{center}
\end{figure}

\begin{figure}
\begin{center}
\includegraphics[height=14em]{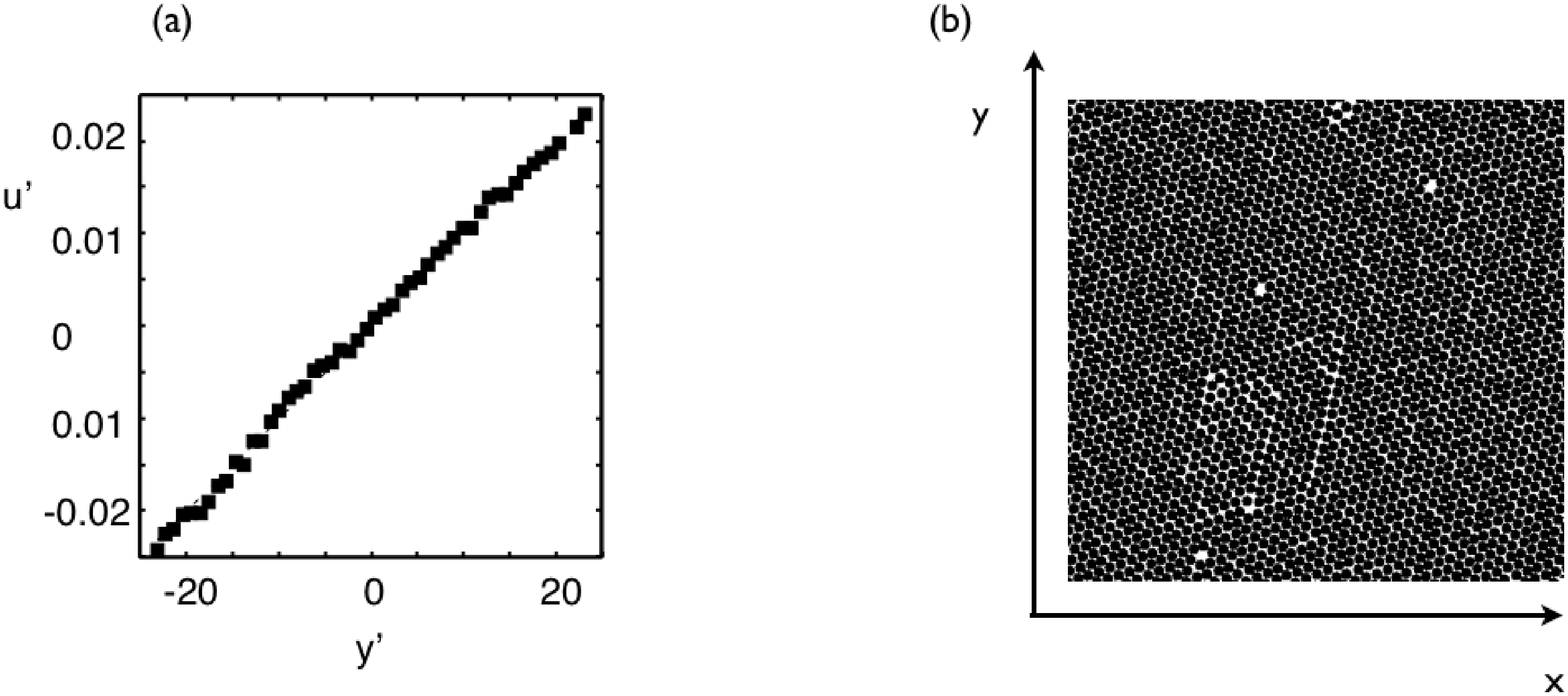}
\caption{ 
  (a) The scaled velocity $u'$ (like in Fig.\ \ref{grad_snap_mono_e0.99})
  for $\phi=0.84$, $\tau_c^* = 1.11 \times 10^{-5}$, and 
  $e=0.999$.
  (b) Snapshot of the {\it monodisperse} system from (a).
% for $\phi=0.76$, $\tau_c^* = 1.11 \times 10^{-5}$, and 
% $e=0.999$.
 }
\label{grad_snap_mono_e0.999}
\end{center}
\end{figure}

\begin{figure}
\begin{center}
\includegraphics[height=14em]{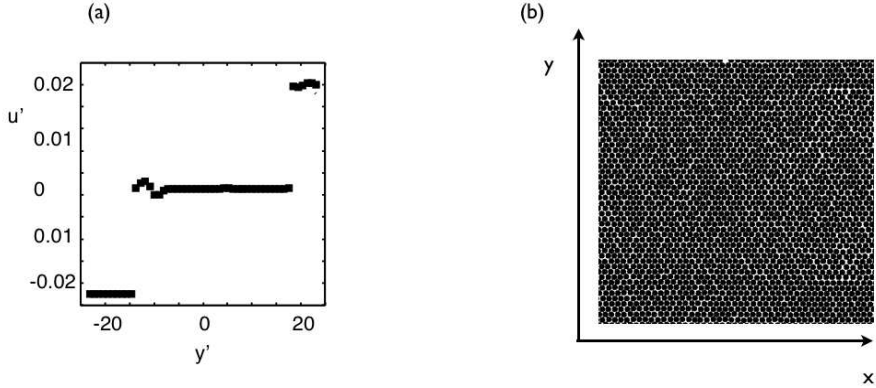}
\caption{ 
  (a) The scaled velocity $u'$ (like in Fig.\ \ref{grad_snap_mono_e0.99})
  for $\phi=0.84$, $\tau_c^* = 1.11 \times 10^{-5}$, and 
  $e=0.9999$.
  (b) Snapshot of the {\it monodisperse} system from (a).
% for $\phi=0.76$, $\tau_c^* = 1.11 \times 10^{-5}$, and 
% $e=0.999$.
 }
\label{grad_snap_mono_e0.9999}
\end{center}
\end{figure}

We have confirmed the existence of shear-bands for $\phi > 0.7$ with $e=0.99$
in Fig.\ \ref{grad_snap_mono_e0.99}.
We plot the velocity $u(y)$ in $x$ direction as a function of $y$
for $\phi=0.76$, $\tau_c^* = 1.11 \times 10^{-5}$, and $e=0.99$ in
Fig.\ \ref{grad_snap_mono_e0.99} (a),
where the velocity gradient exists only in the regions $ y/\sigma_0 < -20$
or  $y/\sigma_0 > 20$.
The apparent inhomogeneity is observed in the snapshot of the system,
see Fig.\ \ref{grad_snap_mono_e0.99}(b).  On the other hand,
such a shear-band could not be observed for the case of $e=0.999$. 
Note that the shear-band formation in our system is different from
that for the dilute case \cite{Tan} in which dense strips align at 45 degrees
relative to the streamwise direction.
%
%YYY How does shear manifest then?  I have no picture in mind since the
%YYY density is so high that the layers of particles should not just slip
%YYY can you pls. clarify (for me - not in the paper)
%
Fig.\ \ref{grad_snap_mono_e0.999} shows that the system is in an uniformly 
sheared state with some density fluctuations, 
see Fig.\ \ref{grad_snap_mono_e0.999}(b).
%WWW NEW:
Actually, here deformations take place irregularly and localized 
-- together with defects and slip planes --
so that the velocity profile looks smooth and linear only
after long-time (or ensemble) averaging.  
For the case of  $e=0.9999$,
almost perfect crystallization is observed, but
slip-lines exist,
%{\bf REMOVE THIS: in the vicinity of the boundary -- they are NOT in the vicinity
%of the boundary!}
see Figs.\ \ref{grad_snap_mono_e0.9999}(a) and (b).

This is in conflict with the observations of 
Ref.\ \citen{Luding09}, where shear-bands were observed
at densities around $\phi \approx 0.70$, $\phi \approx 0.73$, 
and $\phi \approx 0.78$, for $e \ge 0.99$, $e=0.95$, and $e=0.90$,
respectively.
In this paper, for the case of $e=0.999$, no shear band is observed,
however, in the simulation of the sheared inelastically interacting rigid-disks 
with $e=0.998$ in Ref.\ \citen{Luding09}, a shear band was reported.

We identify two differences between the systems in this paper
and Ref.\ \citen{Luding09}.  The first difference is the softness of the disks
that, however, should not affect the results as long as we are close to the rigid-disk 
limit.
%But, from the results of this paper, 
%I think that the softness does not affect the behaviors of the system
%as long as we consider the hard-disk limit.
The second difference is the protocol to obtain a sheared steady state
with density $\phi$.  In this paper, first an equilibrium state with density 
$\phi$ is prepared and then shear flow and dissipation between the particles
is switched on to obtain the sheared steady state.
In contrast, in Ref.\ \citen{Luding09},
the system of sheared inelastically interacting disks was studied by slowly but continuously
increasing the density $\phi$.

%Since the occurrence of the shear band strongly depends on protocols
%to make the sheared state,
%I suppose that the difference about the shear band between this paper
%and  Ref.\ \citen{Luding09} results from the difference of the protocol.

\begin{figure}
\begin{center}
\includegraphics[height=14em]{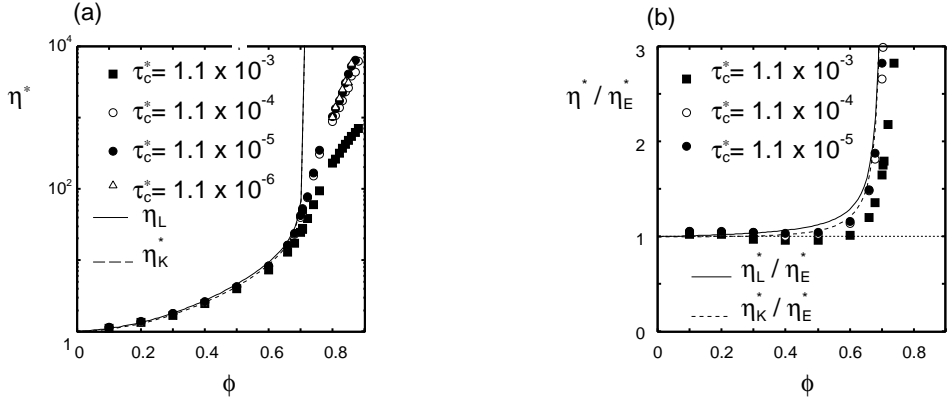}
\caption{ 
  (a) The dimensionless viscosity $\eta^*$ as a function of the area 
  fraction $\phi$ in the {\it monodisperse} system for $e=0.999$, 
  and different $\tau^*_c$ as given in the inset.
% $\tau_c^* = 1.11 \times 10^{-3}$, $1.11 \times 10^{-4}$, 
% and $1.11 \times 10^{-5}$.
  (b) $\eta^*/\eta_E^*$ as a function of the area fraction
  from the same simulations as in (a). 
%{\bf maybe another plot like in my attached Figure ST\_R.ps?
% $\phi$ in the {\it monodisperse} system for $e=0.999$, 
% $\tau_c^* = 1.11 \times 10^{-3},  1.11 \times 10^{-4}, 1.11 \times 10^{-5}$.
%unfortunately the present Fig.\ does not look so convincing ... maybe 
%if you plot the data like I did in Fig.8(b) of my long paper,
%i.e., dividing by the Enskog prediction? and also $\eta_K/\eta_E$
%as well as $\eta_L/\eta_E$?}
}
\label{eta_hard}
\end{center}
\end{figure}

The dimensionless viscosity $\eta^*$ for {\it monodisperse} systems with 
$e=0.999$, and different $\tau^*_c$
%$\tau_c^* = 1.11 \times 10^{-3}$, $1.11 \times 10^{-4}$, 
%and $1.11 \times 10^{-5}$ 
is shown in Fig.\ \ref{eta_hard}.
We note that both $P^*$ and $\eta^*$ converge 
%YYY in the rigid-disk limit
for more rigid disks
$\tau_c^* \to 0$, but not to the empirical expression $\eta_L^*$ 
from Eq.\ \eqref{eta:eq}. It can be used in a wide range of $\phi$, 
as one can see in Fig.\ \ref{eta_hard}(b), 
%$\eta^* / \eta_L^*$ is plotted as a function of $\phi$ for $\phi<0.7$,
%supports the validity of the empirical expression of $\eta_L^*$.
but -- even though behaving qualitatively similar --
the numerical data clearly deviate from $\eta_L^*$: For $\phi > 0.7$,
in the rigid-disk case, $\eta_L^*$ diverges at $\phi_\eta=0.71$, 
whereas $\eta^*$ in the near-rigid case exponentially grows like the
Vogel-Fulcher law, which remains finite above $\phi_\eta$.
%(PLEASE NOTE THAT FIGS. 6(a) and 7(a) SUGGEST THAT THERE IS NO SINGULARITY AROUND $\phi=\phi_\eta$. IT SEEMS THAT THE DIVERGENCE
%OF THE VISCOSITY TAKES PLACE AT THE SAME DIVERGING POINT OF THE PRESSURE, THOUGH THE INVERSE POWER CAN BE A FITTING FUNCTION FOR
%$\phi<\phi_\eta$ ).
%{\bf This is one of the cases where I would like to see $\tau_c$ 
%and $\tau^*_E$ or $\tau^*_\omega$ ... thanks}

\begin{figure}
\begin{center}
\includegraphics[height=14em]{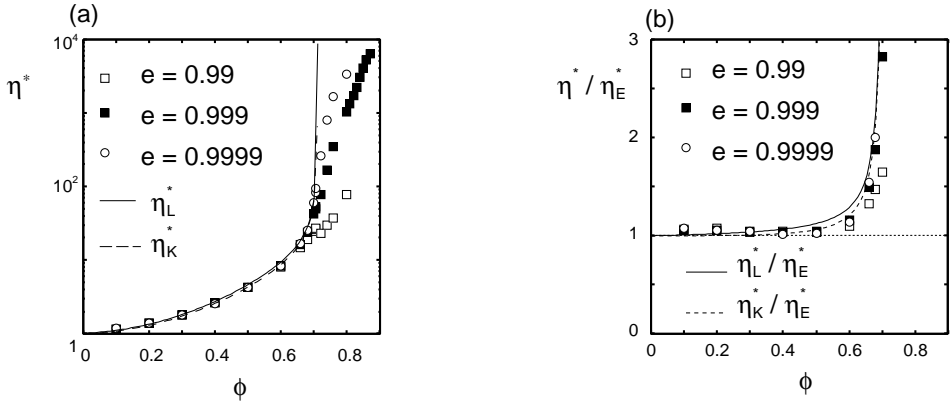}
\caption{ 
  The dimensionless viscosity $\eta^*$ as a function of the area fraction
  $\phi$ in the {\it monodisperse} system for 
  $\tau_c^* = 1.11 \times 10^{-5}$ and $e=0.99$, $0.999$, $0.9999$.
  (b) $\eta^*/\eta_E^*$ as a function of the area fraction from (a).
%{\bf The open square symbols are missing in the right figure - please check.}
 }
\label{eta_elastic}
\end{center}
\end{figure}

%YYY This section, you should check and confirm that you 
%YYY agree with my changes - if not, we should briefly skype again
%YYY because then I have missunderstood something ...
%YYY
The difference between the numerical data for $\eta^*$
 and $\eta^*_L$ results from both elasticity and dissipation,
as shown in Fig.\ \ref{eta_elastic}, where
the dependence of $\eta^*$ on $\phi$ for $\tau_c^* = 1.11 \times 10^{-5}$ 
and different coefficients of restitution $e$ % =0.99, 0.999, 0.9999$ 
are plotted. 
The viscosity $\eta^*$, like the pressure $P^*$, approach 
$\eta^*_L$ and $P^*_Q$ in the elastic limit $e\to1$,
%It is apparent that both $P^*$ and $\eta^*$ in the elastic limit $e\to 1$ 
i.e., they converge to the results of the elastic rigid-disk system.
It should be noted that Figs.\ \ref{eta_hard}(a) and \ref{eta_elastic}(a)
suggest that the singularity around $\phi=\phi_\eta$ is an upper limit,
only realized in the rigid disk limit and for $e \rightarrow 1$. 
As will be discussed below, for given $\tau_c^*$ and $e$, the simulations
deviate more and more from the rigid disk case with increasing density.
The smaller $\tau_c^*$, i.e., the stiffer the disks, the better is the
upper limit approached -- but for finite dissipation and for 
near-rigid disks, there is always a finite density where the elasticity 
(softness) becomes relevant and leads to deviations from the upper limit.  
Above that density, it seems that the divergence of the viscosity takes place
at the same point as the pressure, and another inverse power law can
be a fitting function for $\phi<\phi_\eta$.
%(PLEASE NOTE THAT FIGS. 6(a) and 7(a) SUGGEST THAT THERE IS NO SINGULARITY AROUND $\phi=\phi_\eta$. IT SEEMS THAT THE DIVERGENCE
%OF THE VISCOSITY TAKES PLACE AT THE SAME DIVERGING POINT OF THE PRESSURE, THOUGH THE INVERSE POWER CAN BE A FITTING FUNCTION FOR
%$\phi<\phi_\eta$ ).

\begin{figure}
\begin{center}
\includegraphics[height=14em]{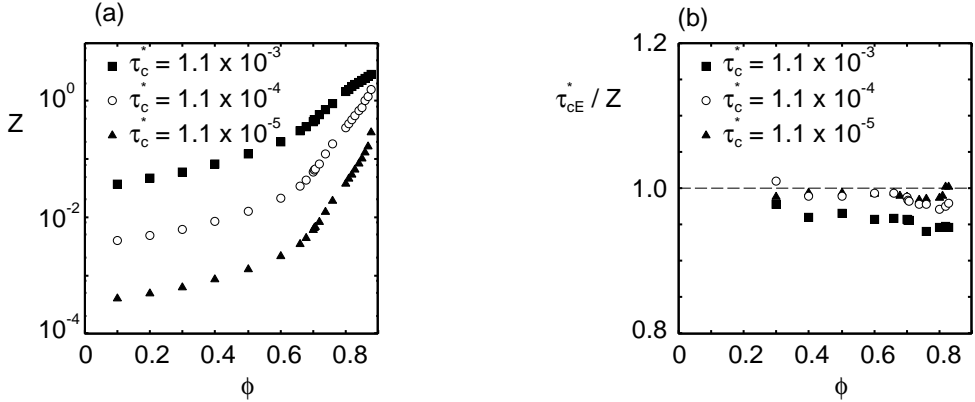}
\caption{ 
  (a) The coordination number $Z$
  plotted as function of the area fraction
  $\phi$ for {\it monodisperse} systems with $e=0.999$ 
  for several $\tau_c^*$ values.
  (b) $\tau_{cE}^* / Z$
%XXX is the figure caption correct, or the figure label?
% Z/tau or tau/Z ???
  plotted as function of the area fraction $\phi$ 
  from the same simulations as in (a).
% XXX Please plot the (b) figure vertical from 0.8 to 1.4 or so
% in order to zoom in to the interesting range.
% for {\it monodisperse} systems with $e=0.999$ 
% for several $\tau_c^*$ values.
% $1.11 \times 10^{-3}$, $1.11 \times 10^{-4}$, and $1.11 \times 10^{-5}$.
%{\bf Please plot this figure with vertical log-axis (see my TC paper
%with McNamara from 1998) and plot lines for the theoretical 
%coordination number $Z = 2C/N \propto t_c/t_E=\tau_c$ 
%(if I remember correctly). This should be true for lower densities
%at least.
%{\bf please set the vertical range from $10^{-4}$ to $10^1$ and move
%the redundant $tau_c^*$ values out from the figure into the caption.  
%Please add a note if the points are really on top of each other or if 
%it only looks like because of the log-scale. Thanks.
%}
}
\label{fig:Z}
\end{center}
\end{figure}

In rigid-disk systems, the coordination number $Z$ 
should be identical to zero because the contacts between 
the particles are instantaneous.
Hence, in the rigid-disk limit of soft-disks,
it is expected that the coordination number $Z$ vanishes,
which is confirmed by Fig.\ \ref{fig:Z}(a).
Here, it should be noted that the coordination number $Z$ is 
almost identical to the dimensionless number $\tau_{cE}^*$ 
\cite{Luding98}. %{\bf proportional or identical?} %XXX
Indeed the relationship $Z\approx \tau_{cE}^*$ 
can be verified in Fig.\ \ref{fig:Z}(b), where we plot
the ratio $\tau_{cE}^* / Z$ as function of the area fraction
$\phi$ for {\it monodisperse} systems with 
$e=0.999$ 
%{\bf Should it be $e=0.999$ -- see caption??? please check?}
and several $\tau_c^*$.
Here, we have measured the coordination number %and $\tau_{cE}^+$
as 
%YYY we should refer to this equation in the appendix
%    and I think from the simulations we do NOT do ensemble 
%    averaging, but just counting - thus we dont need the 
%    brackets here.  True?
\begin{equation}
Z = \sum_i \sum _{j\neq i} \langle \Theta(\sigma_{ij} - r_{ij}) \rangle/N ~.
\label{eq:defZ}
\end{equation}
%XXX old:
% and $\tau_{cE}^* = 2t_c \dot \gamma^2 \eta / \{ (1-e^2) n T\}$, respectively.
% XXX new:
%and $\tau_{cE}^+ = 2 \tau^*_{c\omega}/(1-e^2)$,
% XXX sorry, but if I understand right what you do, this is not true
%            due to the definition of t_E and omega ???
%     we should not say that we plot $\tau_{cE}^*$ here!
%                 = 2 t_c \omega / (1-e^2)
%                 = 2 t_c \dot \gamma^2 \eta / \{ (1-e^2) n T\}$, 
%respectively. The latter is achieved by measuring $\omega = \dot\gamma S/(nT)$,
%according to its definition, and multiplying it by $2 t_c/(1-e^2)$.
%Measuring $\tau_{cE}^*=t_c/t_E \ne \tau_{cE}^+$ directly, 
%by counting the new contacts per unit time,
%leads to somewhat different results close to the jamming transition (data not shown).
%XXX this part is new -^
%{\bf About this was the more recent discussion - It makes a difference if you
%use $\omega$ definition and derive something from it, or if you count the number
%of new contacts ... I hope I have understood correctly and phrased it correctly?}
If we use the mean-field picture, we can understand the relation $Z\approx \tau_{cE}^*$ 
as shown in Appendix \ref{Z:app}.  
%WWW MAYBE NEW: add this:
%The discrepency is probably due to the existence of
%multi-particle contacts.
%
%Let us watch a tracer particle $i$. Introducing a characteristic function $\Theta_{i,j}(t)$ which is
%$\Theta_{i,j}=1$ if the grain $i$ contacts the grain $j$ or $\Theta_{i,j}(t)=0$ otherwise, 
%the coordination number is given by
%XXX WHAT IS THE {1} ?
%$Z\approx \frac{1}{N t_{ob}}\sum_{j\ne i}\int_0^{t_{ob}} dt \Theta_{i,j}(t)$, where $t_{ob}$ is the observed time interval.
%If $t_{ob}$ satisfies $N_t (t_E+t_c)$ with a sufficient large integer $N_t$ and each contact can be regarded as an independent event,
%then we obtain $\int_0^{t_{ob}}\Theta_{i,j}(t)\approx N_t \int_0^{t_E}dt\Theta_{i,j}(t)=N_t t_c$. Substituting this into
%the expression of $Z$ we reach $Z\approx t_c/t_E=\tau_{cE}^*$. Thus, the result in Fig.\ \ref{fig:Z}(b) is reasonable. 

\begin{figure}
\begin{center}
\includegraphics[height=14em]{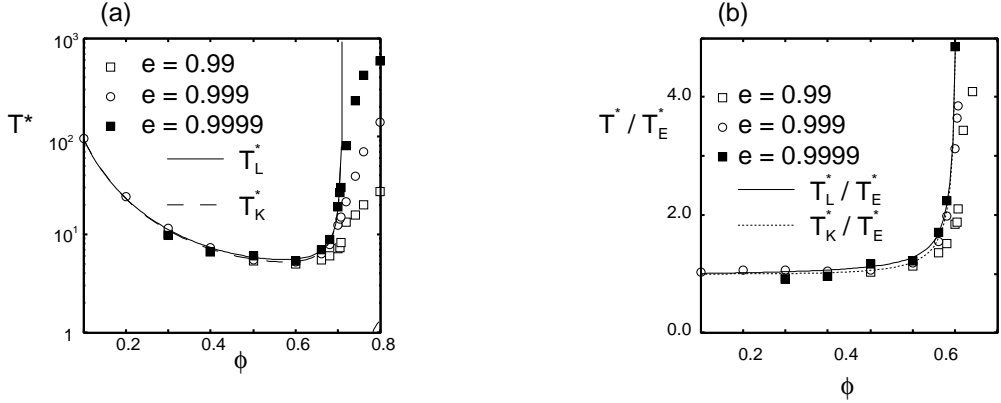}
\caption{ 
  (a) The scaled temperature $T^*$ as a function of the area fraction
  $\phi$ for the {\it monodisperse} system at 
  $\tau_c^* = 1.11 \times 10^{-5}$ and different $e$.
% $e=0.99, 0.999, 0.9999$.
  (b) $T^*/T^*_E$ as a function of the area fraction 
   $\phi$ from the same simulations as in (a).
%  in the {\it monodisperse} system for $\tau_c^* = 1.11 \times {10^5}$
%  and $e=0.99, 0.999, 0.9999$.
%{\bf see my comments for Figs.\ 5 and 6. maybe also plot $T/T_E$?
%-- see my attached PostScript ST\_R.ps.}
\label{T}
}
\end{center}
\end{figure}

We also show the scaled temperature $T^*$ for the soft-sphere
{\it monodisperse} system in Fig.\ \ref{T}.
As expected, $T^*$ approaches the empirical expression $T_K^*$
in Eq.\ \eqref{TK}.  This result also supports our conjecture that
the rigid-disk limit of the soft-disk assemblies coincides with 
the rigid-disk system when the coefficient of restitution $e$
is sufficiently close to unity.
%Thus, all the results support that the behavior of the hard-disk
%limit of soft disks for e<1 coincides with those for hard disks.

\subsection{Poly-disperse systems}
\label{Poly:Subsec}

\begin{figure}
\begin{center}
\includegraphics[height=14em]{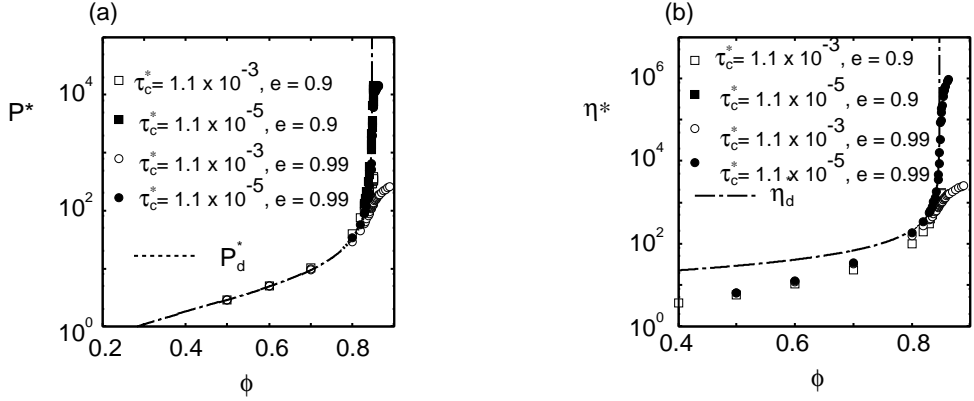}
\caption{ 
  (a) The dimensionless pressure $P^*$ as a function of the area fraction
  $\phi$ for {\it polydisperse} systems with several different $\tau^*_c$ and $e$, 
  where we have used
% $\tau_c^* = 1.11 \times 10^{-3}$, $1.11 \times 10^{-5}$ and $e=0.9$, $0.99$,
  $\phi_{\rm max} = 0.841$ for $P_d^*$ and $\eta^*_d$.
  The prefactor for $P_d^* \propto (\phi_{\rm max} - \phi)^{-1}$
  is chosen as $2\phi_{\rm max}$ \cite{Luding01,Luding02,Luding09}.
%{\bf should we provide the prefactors also? - maybe not in the paper,
%but can you please write them as comments into the figure caption? thanks}
  (b) The dimensionless viscosity $\eta^*$ as a function of the 
  area fraction $\phi$ from the same simulations as in (a).
  Here, we have used the prefactor 7.0 for 
  $\eta_d^* \propto (\phi_{\rm max} - \phi)^{-1}$.
 }
\label{P_eta_poly_wide}
\end{center}
\end{figure}

In order to understand the {\it polydisperse} situation, we also study
systems with different $\tau_c^*$ and different $e$ values -- as in the 
previous subsection.
The reduced pressure $P^*$ and the dimensionless viscosity $\eta^*$
are almost independent of $\tau_c^*$ and $e$ for moderate densities
($\phi<0.8$), as shown in Fig.\ \ref{P_eta_poly_wide},
where $P^*$ and $\eta^*$ are plotted as functions of the 
area fraction $\phi$.
%with $\tau_c^* = 1.11 \times 10^{-3}, 1.11 \times 10^{-5}$ and $e=0.9, 0.99$.
For low densities, the simulation results of $P^*$ agree with 
the scaling given by $P_d^*$,
while the asymptotic scaling behavior of $\eta^*$ is described by $\eta_d^*$ only above $\phi \simeq 0.8$.
Here, we have used $\phi_{\rm max} = 0.841$ for $P^*$ and $\eta^*$ in
Eqs.\ \eqref{Pd:eq} and \eqref{etad:eq}.

\begin{figure}
\begin{center}
\includegraphics[height=14em]{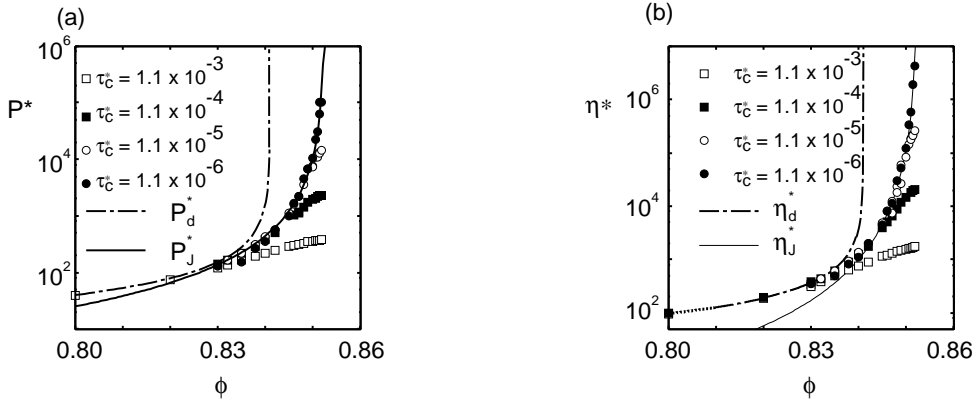}
\caption{ 
  (a) The dimensionless pressure $P^*$ as a function of the area fraction
  $\phi$ for the {\it polydisperse} system for $e=0.9$ and several $\tau^*_c$. % and $0.82<\phi<0.85$.
  (b) The dimensionless viscosity 
  $\eta^*$ as a function of the area fraction
  $\phi$ in the {\it polydisperse} system from the same simulations as those in (a).
% $e=0.9$. % and $0.82<\phi<0.85$.
% We use $b=0.007$ and $\phi_{\rm max}=0.8525$ are used for 
% $P^*_J$ and $\eta^*_J$, and $\phi_\eta = 0.841$ for $P^*_d$ and $\eta^*_d$. 
   Here, we have used $\phi_{J}=0.8525$ for $P^*_J$ and $\eta^*_J$,
   and $\phi_{\rm max} = 0.841$ for $P^*_d$ and $\eta^*_d$.
  The prefactors for $P_d^* \propto (\phi_{\rm max} - \phi)^{-1}$
  and $\eta_d^* \propto (\phi_{\rm max} - \phi)^{-1}$
  are $2\phi_{\rm max}$ and $7.0$, respectively.
  For $P_J^* \propto (\phi_{J} - \phi)^{-2}$
  and $\eta_J^* \propto (\phi_{J} - \phi)^{-3}$,
  the prefactors are chosen as $0.07$ and $0.002$, respectively.
%YYY in the left figure, the 'star' at P_J^* is different from P_d^*
%    -- only a tiny cosmetic change --
%YYY Here the real question:
%  In Fig.10-left, the data seem to collapse with P_d^*, but when 
%  zooming in in Fig.11-left, they are further right - probably this
%  is not visible?  Just want to make sure ...
%
 }
\label{P_eta_poly_e0.9}
\end{center}
\end{figure}

However, when looking more closely, there are distinct differences
between $P^*$ and $P_d^*$, and between $\eta^*$ and $\eta^*_d$ for 
$\phi > 0.83$.  In Fig.\ \ref{P_eta_poly_e0.9}, 
$P^*$ and $\eta^*$ are plotted from {\it polydisperse} systems with 
rather strong dissipation, $e=0.9$,
where we have used particular values for $\phi_{\rm max} = 0.841$ 
and $\phi_J = 0.8525$ in order to visualize their different behavior.
Although $P^*$ is still finite for $\phi > \phi_{\rm max}$ in the
hard disk limit, even for the smallest $\tau^*_c$ values,
both $P_d^*$ and $\eta_d^*$ diverge at $\phi_{\rm max}$ as 
$(\phi_{\rm max} - \phi)^{-1}$.
On the other hand, in the same high density range,
$P^*$ and $\eta^*$ are consistent with $P_J^*$ \eqref{scaling:P} and 
$\eta_J^*$ \eqref{scaling:eta}~\cite{Otsuki:PTP,Otsuki:PRE}
in the rigid-disk limit ($\tau_c^* = 1.11 \times 10^{-6}$),
as will be shown below.

\begin{figure}
\begin{center}
\includegraphics[height=14em]{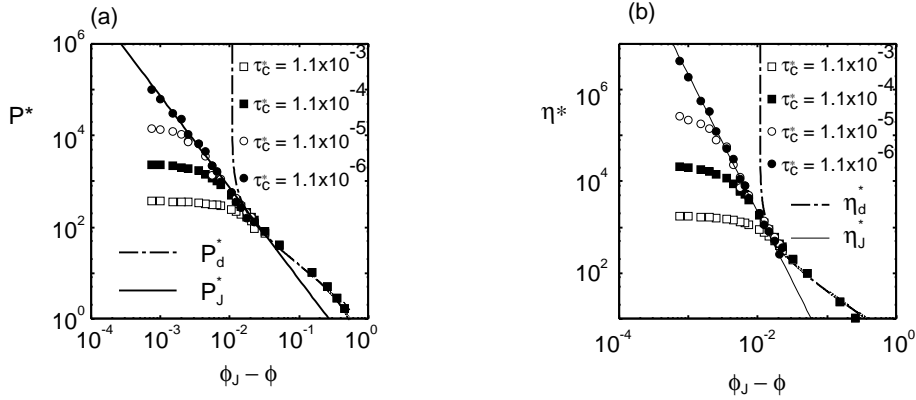}
\caption{ 
  (a) The reduced pressure $P^*$ plotted as a function of 
  $\phi_J - \phi$ for {\it polydisperse} systems with $e=0.9$
  and several $\tau^*_c$ based on 
  the simulations used for Fig.\ \ref{P_eta_poly_e0.9}.
  (b) The dimensionless viscosity $\eta^*$ from the same simulations 
  as those in (a).
% We use $b=0.007$ and $\phi_{\rm max}=0.8525$ are used 
% for $P^*_J$ and $\eta^*_J$, and $\phi_\eta = 0.841$ for 
% $P^*_d$ and $\eta^*_d$. 
  Here, we have used $\phi_{J}=0.8525$ for $P^*_J$ and $\eta^*_J$,
  and $\phi_{\rm max} = 0.841$ for $P^*_d$ and $\eta^*_d$.
 }
\label{P_eta_poly_log_e0.9}
\end{center}
\end{figure}

In order to verify whether the critical behavior of $P^*$ and $\eta^*$
can be described by $P_J^*$ and $\eta_J^*$, 
we plot $P^*$ and $\eta^*$ as functions of $\phi_J - \phi$ 
in Fig.\ \ref{P_eta_poly_log_e0.9}.
Here, we plot only the data for $\phi < \phi_J$
because we discuss the scaling behavior of $P^*$ and $\eta^*$
in the unjammed regime in this paper.
$P^*$ and $\eta^*$ in the rigid-disk limit 
approach $P_J^*$ and $\eta_J^*$, which satisfy 
$(\phi_J - \phi)^{-2}$ and $(\phi_J - \phi)^{-3}$, 
respectively.

\begin{figure}
\begin{center}
\includegraphics[height=14em]{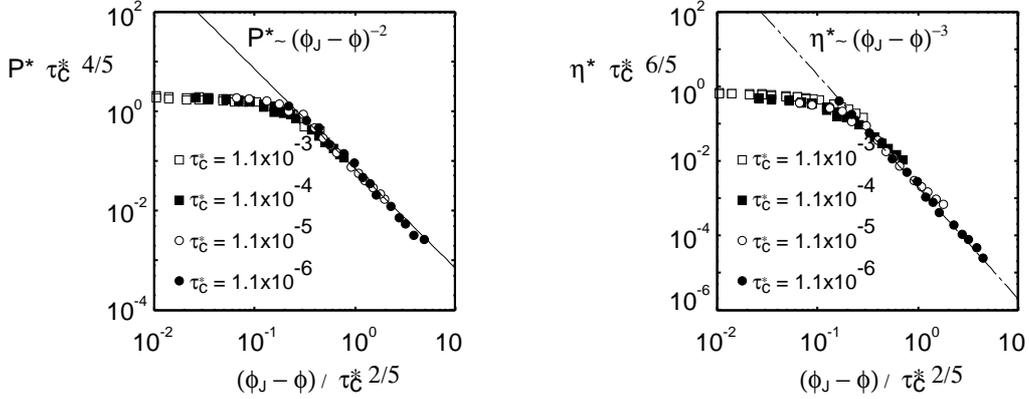}
\caption{ 
  (a) Plots of $P^* \tau_c^{*4/5}$ versus 
 $(\phi_J-\phi)/\tau_c^{*2/5}$
  for {\it polydisperse} systems with $e=0.9$
  and several $\tau^*_c$.
  (b) Plots of $\eta^* \tau_c^{*6/5}$ versus 
 $(\phi_J-\phi)/\tau_c^{*2/5}$
  for {\it polydisperse} systems with $e=0.9$
  and several $\tau^*_c$.
 }
\label{scale_P_eta}
\end{center}
\end{figure}

It should be noted that the plateaus in Fig.\ \ref{P_eta_poly_log_e0.9},
close to the jamming transition point, for $\phi \simeq \phi_J$, 
can also be predicted from the scaling theory, by rewriting
Eqs.\ \eqref{scaling}--\eqref{exponents}. More specifically,
the arguments are taken to the power $-1/\alpha$:
% and for the prefactor
%the relation $\dot \gamma^{1/\alpha} \propto |\Phi|$ is used:
%WWW ??? what is done here ??? can it be said in simple words ???
%WWW - rewritten a little - The scaling relations
%in Eqs.\ \eqref{scaling}--\eqref{exponents} are rewritten as
\begin{equation}
T  =  \dot \gamma^{x_{\Phi}/\alpha} {\cal T}'_{\pm}
\left(\frac{|\Phi|}{\dot\gamma^{1/\alpha}}\right),  \
S  =  \dot \gamma^{y_{\Phi}/\alpha}{\cal S}'_{\pm}
\left(\frac{|\Phi|}{\dot\gamma^{1/\alpha}}\right),  \
P  =  \dot \gamma^{y_{\Phi}'/\alpha}{\cal P}'_{\pm}
\left(\frac{|\Phi|}{\dot\gamma^{1/\alpha}}\right), 
\end{equation}
where we have introduced
${\cal T}'_{\pm}(x) = x^{-x_\Phi}{\cal T}_{\pm}(x^{-\alpha})$,
${\cal S}'_{\pm}(x) = x^{-y_\Phi}{\cal S}_{\pm}(x^{-\alpha})$, and
${\cal P}'_{\pm}(x) = x^{-y'_\Phi}{\cal P}_{\pm}(x^{-\alpha})$.
The scaling functions satisfy
$\lim_{x\to 0}{\cal T}'_{\pm}(x)=\lim_{x\to 0}{\cal S}'_{\pm}(x)=\lim_{x\to 0}{\cal P}'_{\pm}(x) = const.$
%Since temperature, shear stress and pressure 
%behave as $T \sim \dot \gamma ^{x_\Phi/\alpha}$, 
%$S \sim \dot \gamma^{y_\Phi/\alpha}$ and 
%$P \sim \dot \gamma^{y_\Phi'/\alpha}$,
Substituting these relations into Eqs.\ \eqref{P:def}
\eqref{eta:def}, with Eqs.\ \eqref{exponents}, $\eta=S/\dot\gamma$,
$\Delta=1$, and the definition of $\tau_c^*$ given by Eq.\ \eqref{tau:def}, 
the scaling relations of $P^*$ and $\eta^*$ are obtained as
\begin{equation}
P^*  =  \tau_c^{*-4/5} {\cal P}^*_{\pm}
\left(\frac{|\Phi|}{\tau_c^{*2/5}}\right),  \
\eta^*  =  \tau_c^{*-6/5} {\cal H}^*_{\pm}
\left(\frac{|\Phi|}{\tau_c^{*2/5}}\right).
\end{equation}
Here, the scaling functions satisfy 
$\lim_{x\to 0}{\cal P}^*_{\pm}(x)=\lim_{x\to 0}{\cal H}^*_{\pm}(x) = const.$
 Therefore,
 the plateau for $P^*$ and $\eta^*$ in Fig.\ \ref{P_eta_poly_log_e0.9}
should be  proportional to $(1/\tau_c^*)^{4/5}$ and $(1/\tau_c^*)^{6/5}$,
respectively, which is confirmed by Fig.\ \ref{scale_P_eta},
where we plot $P^* \tau_c^{*4/5}$ and $\eta^* \tau_c^{*6/5}$
as a function of $(\phi_J-\phi)/\tau_c^{*2/5}$.
%Since temperature, shear stress and pressure 
%the constant values of $P^*$ and $\eta^*$ 
%are expected to be proportional to $P^* \sim \dot \gamma^{-4/5}$ 
%and $\eta^* \sim \dot \gamma^{-6/5}$.
%Using the definition of $\tau_c^*$ given by Eq.\ \eqref{tau:def},
% the plateau for $P^*$ and $\eta^*$ in Fig.\ \ref{P_eta_poly_log_e0.9}
%should be  proportional to $(1/\tau_c^*)^{4/5}$ and $(1/\tau_c^*)^{6/5}$,
%respectively, which is confirmed by Fig.\ \ref{scale_P_eta},
%where we plot $P^* \tau_c^{*4/5}$ and $\eta^* \tau_c^{*6/5}$
%as a function of $(\phi_J-\phi)/\tau_c^{*2/5}$.
%YYY I am not sure that I understand the argument - can you please detail a little more here?
%    I think I could figure out the scaling of the vertical axis, but not that of the horizontal axis.
%
%(not shown). {\bf Have you shown this in another paper?
%If yes, pls. add the reference, if not, maybe its worth showing it here?}

\begin{figure}
\begin{center}
\includegraphics[height=14em]{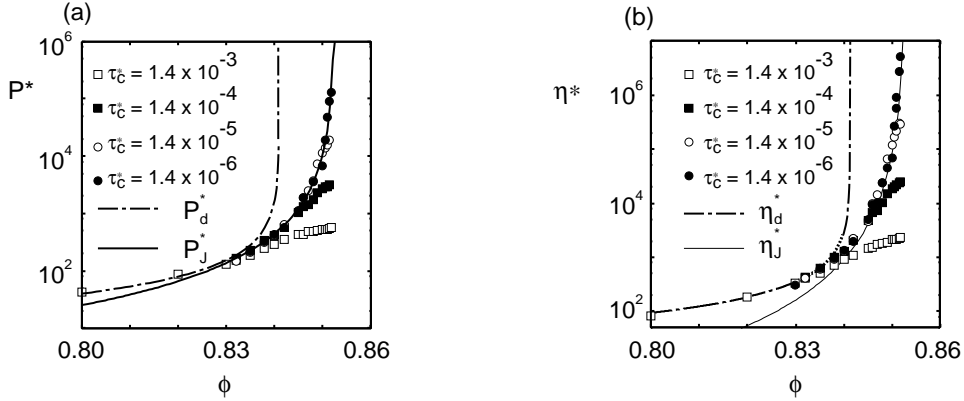}
\caption{ 
  (a) The reduced pressure $P^*$ as a function of the area fraction
  $\phi$ for the {\it polydisperse} system with $e=0.1$ and several
  $\tau^*_c$.
  (b) The dimensionless viscosity $\eta^*$ from the same data as those in (a).
  We have used $b=0.07$ and $\phi_J=0.8525$ for $P_J$ and $\eta_J$,
  and $\phi_{\rm max} = 0.841$ for $P_d$ and $\eta_d$. 
   We used $\phi_{J}=0.8525$ for $P^*_J$ and $\eta^*_J$,
   and $\phi_{\rm max} = 0.841$ for $P^*_d$ and $\eta^*_d$
  The prefactors for $P_d^* \propto (\phi_{\rm max} - \phi)^{-1}$
  and $\eta_d^* \propto (\phi_{\rm max} - \phi)^{-1}$
  are $2\phi_{\rm max}$ and $7.0$, respectively.
  The prefactors for $P_J^* \propto (\phi_{J} - \phi)^{-2}$
  and $\eta_J^* \propto (\phi_{J} - \phi)^{-3}$
   are given by $0.07$ and $0.002$, respectively.
 }
\label{P_eta_poly_e0.1}
\end{center}
\end{figure}

\begin{figure}
\begin{center}
\includegraphics[height=14em]{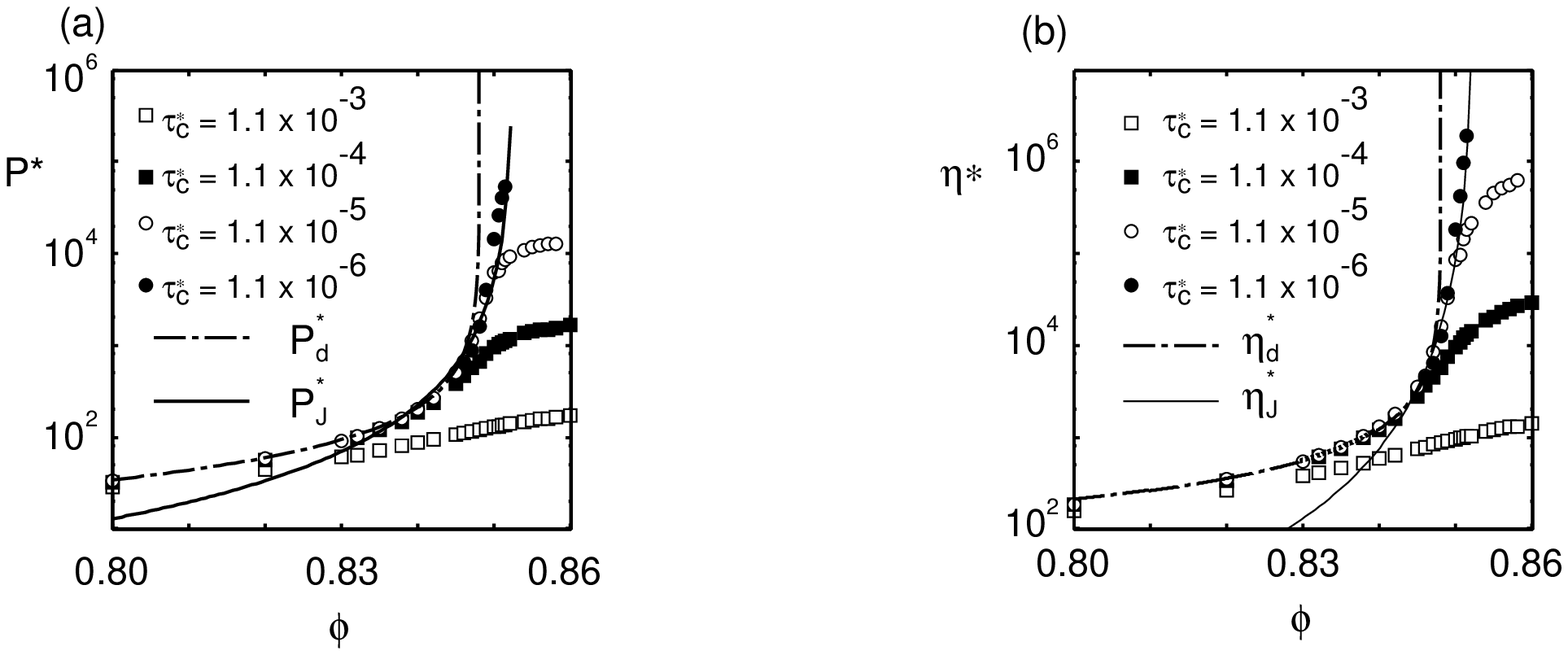}
\caption{ 
  (a) The reduced pressure $P^*$ as a function of the area fraction
  $\phi$ for the {\it polydisperse} system with $e=0.99$  and several
  $\tau^*_c$.
  (b) The dimensionless viscosity $\eta^*$ obtained from the same data as those in (a).
  We have used $\phi_J=0.8525$ for $P_J$ and $\eta_J$,
  and $\phi_{\rm max} = 0.848$ for $P_d$ and $\eta_d$. 
  The prefactors for $P_d^* \propto (\phi_{\rm max} - \phi)^{-1}$
  and $\eta_d^* \propto (\phi_{\rm max} - \phi)^{-1}$
  are $2\phi_{\rm max}$ and $10.0$, respectively.
  The prefactors for $P_J^* \propto (\phi_{J} - \phi)^{-2}$
  and $\eta_J^* \propto (\phi_{J} - \phi)^{-3}$,
   are $0.035$ and $0.0015$, respectively.
 }
\label{P_eta_poly_e0.99}
\end{center}
\end{figure}

\begin{figure}
\begin{center}
\includegraphics[height=14em]{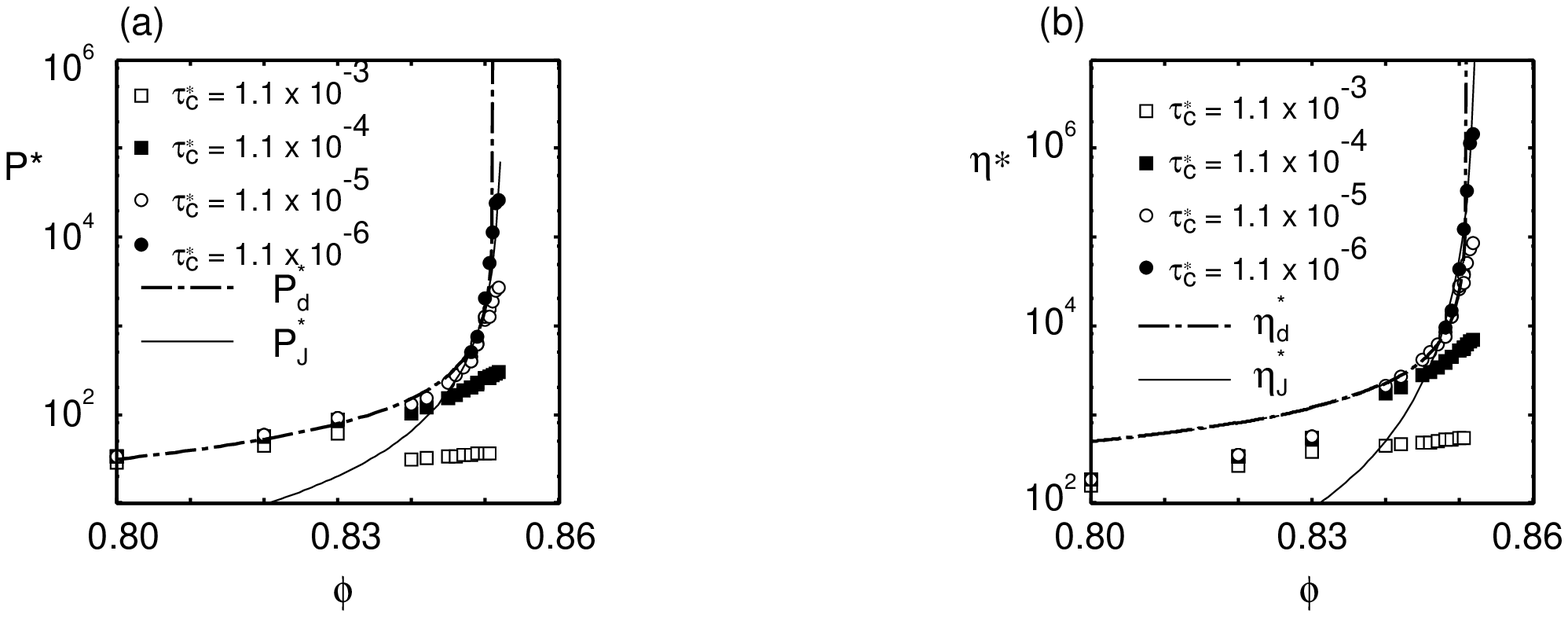}
\caption{ 
  (a) The reduced pressure $P^*$ as a function of the area fraction
  $\phi$ for the {\it polydisperse} system for $e=0.998$ and several
  $\tau^*_c$.
  (b) The dimensionless viscosity $\eta^*$ obtained from the same data as those in (a).
  We have used $\phi_J=0.8525$ for $P^*_J$ and $\eta_J$,
  and $\phi_{\rm max} = 0.851$ for $P^*_d$ and $\eta_d$. 
  The prefactors for $P_d^* \propto (\phi_{\rm max} - \phi)^{-1}$
  and $\eta_d^* \propto (\phi_{\rm max} - \phi)^{-1}$
  are $2\phi_{\rm max}$ and $25.0$, respectively.
  The prefactors for $P_J^* \propto (\phi_{J} - \phi)^{-2}$
  and $\eta_J^* \propto (\phi_{J} - \phi)^{-3}$,
  are $0.01$ and $0.001$, respectively.
 }
\label{P_eta_poly_e0.998}
\end{center}
\end{figure}

Whether the simulation pressure is described by $P_d^*$ or $P_J^*$, 
and whether the viscosity is given by $\eta_d^*$ or $\eta_J^*$, 
strongly depends on the coefficient of restitution $e$.
In Figs.\ \ref{P_eta_poly_e0.1}--\ref{P_eta_poly_e0.998},
we plot $P^*$ and $\eta^*$ as functions of $\phi$
for various $e$, involving the very high dissipation case $e=0.1$, 
an intermediate case $e=0.99$, and a low dissipation case $e=0.998$.
%{\bf Is it $e=0.999$ as written here or $e=0.998$ as 
%written below?}
%
Using fitting values $\phi_{\rm max}=0.841$, $0.848$, and $0.851$, 
based on a fit starting from very low densities, %YYY True???
corresponding to various $e=0.1$, $0.99$ and $0.998$, respectively,
we can approximate the data of $P^*$ best by
$P_d^* = 2 \phi_{\rm max}/(\phi_{\rm max} - \phi)$.
%{\bf ??? I don't understand - HOW are they estimated? From a low
%density fit? It is clear that the d-predictions are wrong if the
%divergence density is wrong. ... please detail here ... thanks.}
On the other hand, we assume that $\phi_J$ is independent of $e$, 
and fix $\phi_J=0.8525$ for all $e$,
as confirmed this by our numerical simulations. 
%YYY -- I dont understand the following sentence - it is not clear in this context
%Indeed, the critical density is only determined by
%$\tau_{cE}^*$ in the hard disk limit, where the inelasticity or $\zeta$
%is not important for $t_c$ for large $k$.

\begin{figure}
\begin{center}
\includegraphics[height=14em]{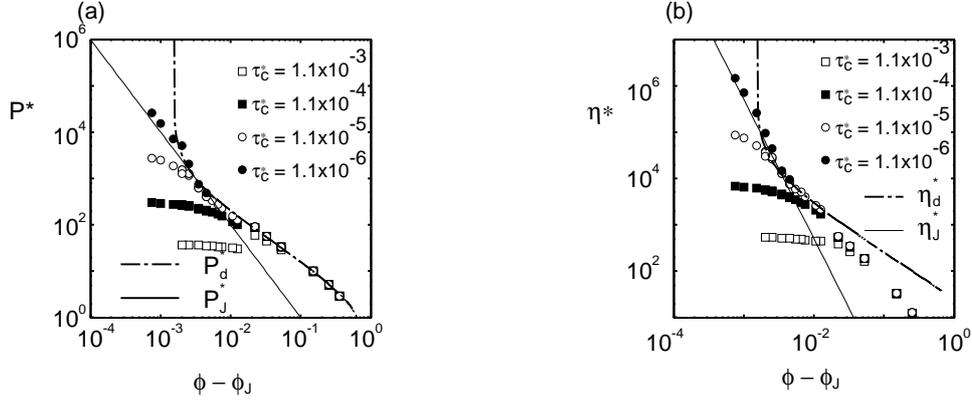}
\caption{ 
  (a) The reduced pressure $P^*$ plotted as a function of 
  $\phi_J - \phi$ for {\it polydisperse} systems with $e=0.998$
  and several $\tau^*_c$ based on 
  the simulations used for Fig.\ \ref{P_eta_poly_e0.998}.
  (b) The dimensionless viscosity $\eta^*$ obtained from the same simulations 
  as those in (a).
  We have used $\phi_J=0.8525$ for $P^*_J$ and $\eta_J$,
  and $\phi_{\rm max} = 0.851$ for $P^*_d$ and $\eta_d$. 
  The prefactors for $P_d^* \propto (\phi_{\rm max} - \phi)^{-1}$
  and $\eta_d^* \propto (\phi_{\rm max} - \phi)^{-1}$
  are $2\phi_{\rm max}$ and $25.0$, respectively.
  The prefactors for $P_J^* \propto (\phi_{J} - \phi)^{-2}$
  and $\eta_J^* \propto (\phi_{J} - \phi)^{-3}$,
  are $0.01$ and $0.001$, respectively.
 }
\label{P_eta_poly_log_e0.998}
\end{center}
\end{figure}

%
Even in the case of strong inelasticity ($e=0.1$), as
shown in Fig.\ \ref{P_eta_poly_e0.1},
$P_J^*$ and $\eta_J^*$ characterize the behavior of $P^*$ and $\eta^*$ 
near the jamming transition point, while $P_d^*$ and $\eta_d^*$ deviate
for $\phi > 0.83$.  The range where $P_J^*$ and $\eta_J^*$
characterize the pressure and the viscosity becomes narrower as 
$e \to 1$, while the range of validity of $P_d^*$ becomes wider,
as shown in Figs.\ \ref{P_eta_poly_e0.99} and \ref{P_eta_poly_e0.998}.
For $e=0.998$ (Fig.\ \ref{P_eta_poly_e0.998}),
the difference between $P_d^*$ and $P_J^*$ appears only 
in a small region of $\phi$ which is shown in Fig.\ \ref{P_eta_poly_log_e0.998}.

Since the scaling behaviors of $P^*$ and $\eta^*$ 
agree with $P_J^*$ and $\eta_J^*$ near $\phi_J$, 
we conclude that the critical behavior for 
%YYY introduced 'near-rigid' below  -- true?
inelastic near-rigid systems is well described by $P_J^*$ and $\eta_J^*$, 
as proposed in Refs.~\citen{Otsuki:PTP,Otsuki:PRE}.
The scaling plot in Fig.\ \ref{scale_P_eta}
supports the validity of the critical behaviors
concerning both the plateaus and the lower densities.
%WWW slightly changed above
However, such predictions cannot be used for almost elastic 
and perfectly elastic systems, neither mono- or polydisperse,
whose critical behavior is described by $P_d^*$ and $\eta_d^*$
instead.
%YYY here one sentence about the scaling of the plateaus - i.e.
%the good quality thereof -- is missing. I did not succeed to write it.

%{\bf For the last figure\ \ref{P_eta_poly_e0.998}, I cannot see the difference
%between the d- and the J-curves anymore - only a log-log plot like Fig.11
%could show this ... may I see that?}

%{\bf And another thing: For low dissipation, we will have very high 
%temperature and thus very small time between the collisions ... therefore,
%I expect $\tau_c$ (AND $\tau_E^*$) to be VERY large (VERY small)
%respectively. I have not yet figured out what that means ... but it
%seems to indicate that in the hard-disk limit and in 
%this regimes of $\tau_c$ (AND $\tau_E^*$),
%the d-values hold ... see other discussion above.}
%
%{\bf Michio : As I have shown in the last mail,
%$\tau_\omega$ can be regarded as the criterion for the change of scaling laws.\\
%Stefan:\\
%But unfortunately, that is not very well described in this section.
%}

\subsection{Dimensionless numbers and a criterion for the two scaling regimes}
\label{Time:Subsec}

\begin{figure}
\begin{center}
\includegraphics[height=14em]{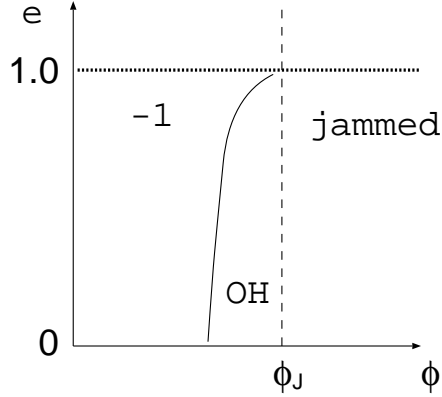}
\caption{ 
A schematic phase diagram of the region (-1) satisfying
Eqs.\ (\ref{Pd:eq}), (\ref{etad:eq}) and 
the region (OH) satisfying 
Eqs.\ (\ref{scaling:P}), (\ref{scaling:eta}).
 }
\label{phase}
\end{center}
\end{figure}

In Sec.\ \ref{Poly:Subsec}, we reported a crossover
from the region satisfying Eqs.\ (\ref{Pd:eq}) and (\ref{etad:eq})
to the region satisfying Eqs.\ (\ref{scaling:P}) and (\ref{scaling:eta}).
Figure \ref{phase} presents a schematic phase diagram 
in the plane of the restitution coefficient $e$ and the area fraction $\phi$,
where -1 denotes the region satisfying the scaling relations given by Eqs.\ (\ref{Pd:eq}) and (\ref{etad:eq}), and 
OH denotes the region satisfying the scalings given by Eqs.\ (\ref{scaling:P}) and (\ref{scaling:eta}).
For each $e$, the high density region satisfies Eqs.\ (\ref{scaling:P}) and (\ref{scaling:eta}), 
while the low density region satisfies the scalings given by Eqs.\ (\ref{Pd:eq}) and (\ref{etad:eq}).
As the restitution coefficient approaches unity, the region of 
OH becomes ``narrower'', and disappears in the elastic limit.

%One may suppose that the criterion to distinguish the two regions 
%is given by a dimensionless number $X$ as
%\begin{enumerate}
%\item $X \gg A$  gives the scalings (\ref{scaling:P}) and (\ref{scaling:eta})
%\item $X \ll A$ gives  the scalings (\ref{Pd:eq}) and (\ref{etad:eq})
%\end{enumerate}
%where $A$ is a constant,
%or the opposite case
%\begin{enumerate}
%\item $X \ll A$  gives the scalings (\ref{scaling:P}) and (\ref{scaling:eta})
%\item $X \gg A$ gives  the scalings (\ref{Pd:eq}) and (\ref{etad:eq}).
%\end{enumerate}

Now, let us discuss which of the dimensionless numbers
$\tau_E^*$, $\tau_{cE}^*$, $\tau_\omega^*$ or $\tau_{c\omega}^{*}$
can be used as the criterion to distinguish between the two scaling regimes.
It should be noted that the dimensionless number for the criterion must be a monotonic function of $\phi$, 
because the scaling relations Eqs.\ (\ref{Pd:eq}) and (\ref{etad:eq}) appear in the higher density region
and the scaling relations Eqs.\ (\ref{scaling:P}) and (\ref{scaling:eta}) appear in the lower density region
regardless to other parameters.
%XXX -- in the hard disk limit. 

\begin{figure}
\begin{center}
\includegraphics[height=14em]{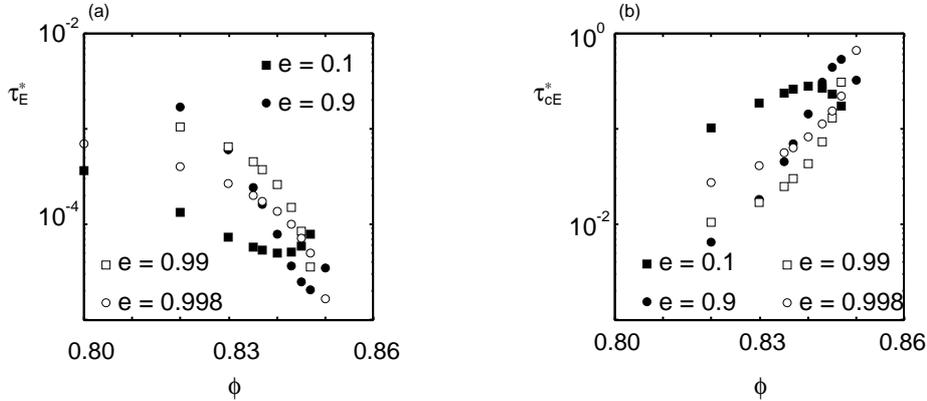}
\caption{ 
(a)  $\phi$ dependence on $\tau_{E}^*$ and
(b) $\phi$ dependence on $\tau_{cE}^*$, for various $e$,
from simulations with $\tau_c^*=1.1 \times 10^{-5}$.
}
\label{tauE_poly_elastic}
\end{center}
\end{figure}

First, let us consider $\tau_E^*$.
We expect that $\tau_E^* < A$ or $\tau_E^* > A$ is the criterion
for the scaling regime given by (\ref{scaling:P}) and (\ref{scaling:eta}),
where $A$ is a constant.
However, since 
$\tau_E^*$ is not a monotonic function of the area fraction $\phi$ 
and the restitution coefficient $e$, 
as shown in Fig.\ \ref{tauE_poly_elastic}(a),
we conclude that neither $\tau_E^* < A$ or $\tau_E^* > A$  is appropriate
for the criterion.

Similar to the case of $\tau_{E}^*$, $\tau_{cE}^*$
in not a monotonic function of $\phi$ and $e$, as shown in 
Fig.\ \ref{tauE_poly_elastic}(b).
Therefore,
we conclude that $\tau_{cE}^*$
is not an appropriate dimensionless time for the criterion.

%Since Eqs.\ (\ref{scaling:P}) and (\ref{scaling:eta})
%are satisfied in the high density region and
%$\tau_E^*$ is a decreasing function of the density, 
%which is shown in Fig.\ \ref{tauE_poly_elastic},
%the condition of $\tau_E^*<A$ is the possible criterion, where $A$ is a constant.
%between
%$\tau_E^*>A$ or $\tau_E^*<A$.
%Unfortunately, 
%this condition is not appropriate to characterize 
%the condition of the jamming scaling. 
%Indeed, as shown in Fig.\ \ref{tauE_poly_elastic}(a), 
%$\tau_E^*$ decreases as the restitution constant $e$ increases.
%This means that the region satisfying $\tau_E^*<A$ is broader as the restitution constant increases, which is opposite to the numerical observation.
%Thus, we conclude that $\tau_E^*$ is not an appropriate dimensionless number for the criterion.
%Similarly, $\tau_s^*$ is not appropriate, because $\tau_s^*$ is identical to $\tau_E^*$ in the hard disk limit.

%Second, let us consider $\tau_{cE}^*$. 
%As shown in Fig.\ \ref{tauE_poly_elastic}(b), $\tau_{cE}^*$ is
%an increasing function of $\phi$.
%Thus, from similar consideration as in the case of $\tau_E^*$,
%between $\tau_{cE}^* > A$ and $\tau_{cE}^* < A$,
%$\tau_{cE}^* > A$ is the only the possible condition for the scaling
%regime given by Eqs.\ (\ref{scaling:P}) and (\ref{scaling:eta}).
%However, this condition is not appropriate.
%$\tau_{cE}^*$ is an increasing function of $e$,
%as shown in Fig.\ \ref{tauE_poly_elastic}(b),
%which means that the region satisfying
%$\tau_{cE}^* > A$ is broader as the restitution constant increases.
%This also means that the region satisfying 
%Eqs.\ (\ref{scaling:P}) and (\ref{scaling:eta})
%is broader as $e\to 1$,
%which is opposite to the numerical observation.
%Thus, we conclude that $\tau_{cE}^*$ is not an appropriate dimensionless time
%for the criterion.

\begin{figure}
\begin{center}
\includegraphics[height=14em]{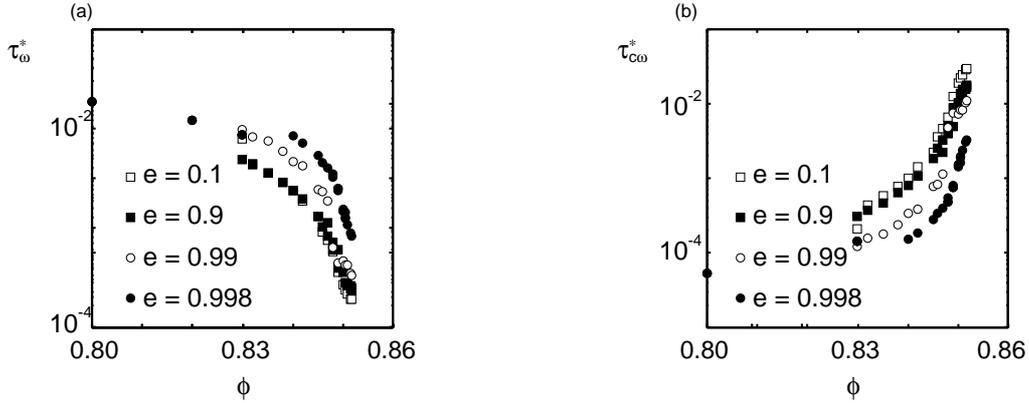}
\caption{ 
(a) $\phi$ dependence on $\tau_{\omega}^*$ and
(b) $\phi$ dependence on $\tau_{c\omega}^{*}$ for various $e$,
from simulations with $\tau_c^*=1.1 \times 10^{-5}$.
%XXX
 }
\label{tauw_poly_elastic}
\end{center}
\end{figure}

Finally, let us consider $\tau_{\omega}^*$ and $\tau_{c\omega}^*$,
which are respectively related with $\tau_E^*$ and $\tau_{cE}^*$ 
as $\tau_\omega^* \approx 2\tau_{E}^* / (1-e^2)$
and $\tau_{c\omega}^{*}\approx (1-e^2)\tau_{cE}^*/2$
in the collisional regime, but their dependency on $\phi$ and $e$ 
differs from those of $\tau_E^*$ and $\tau_{cE}^*$, as shown in
Figs.\ \ref{tauw_poly_elastic}(a)  and \ref{tauw_poly_elastic}(b).
Both $\tau_{\omega}^*$ and $\tau_{c\omega}^*$ are
monotonic functions of  $\phi$ and $e$.
Since Eqs.\ (\ref{scaling:P}) and (\ref{scaling:eta})
are satisfied in the high density region and
$\tau_\omega^*$ and $\tau_{c\omega}^*$ are respectively decreasing
and increasing functions of the density $\phi$,
$\tau_\omega^*<A$ and $\tau_{c\omega}^*>A$ are the possible conditions 
for the scaling given by Eqs.\ (\ref{scaling:P}) and (\ref{scaling:eta}).
These conditions are also consistent with 
the dependencies of $\tau_\omega^*$ and $\tau_{c\omega}^*$ on $e$.
Indeed, $\tau_\omega^*$ increases as the restitution constant increases,
and $\tau_{c\omega}^{*}$ is a decreasing function of $e$.
This means that the regions satisfying 
$\tau_\omega^*<A$ and $\tau_{c\omega}^*>A$
are narrower as the restitution constant increases, 
which is consistent with the numerical observation.
Therefore, $\tau_\omega^*<A$ and $\tau_{c\omega}^*>A$ 
are the only two possible candidates to characterize the system with
respect to their scaling behavior.
It should be noted that $\tau_{c\omega}^*$  tends to zero in the 
hard disk limit $\tau_c^* \to 0$. In this sense,
to use $\tau_{c\omega}^*$ might involve a conceptual difficulty, 
even though $\tau_{c\omega}^*$ is finite in the jamming region. 

\section{Conclusion and Discussion}
\label{Conclusion:Sec}

In conclusion, we have investigated the dimensionless pressure
$P^*$ and the dimensionless viscosity $\eta^*$ of two-dimensional 
soft disk systems and have payed special attention to the rigid-disk 
limit of inelastically interacting systems, while near-rigid disks still 
have some elasticity (``softness'').

For {\it monodisperse} systems, as the system approaches the elastic 
limit, $e \to 1$, both $P^*$ and $\eta^*$ for $\phi<\phi_\eta=0.71$ 
approach the results of 
elastic rigid-disk systems, where the viscosity increases rapidly 
around $\phi=\phi_\eta$ due to ordering (crystallization) effects,
while the %XXX viscosity 
pressure for $\phi>\phi_\eta$ is still finite \cite{Garcia}.
This result is consistent with Ref.\ \citen{Mitarai},
where Mitarai and Nakanishi suggested
that the behavior of soft-disks in dilute collisional flow
converges to that of rigid-disks in the rigid-disk limit. 
%HOWEVER, THE VISCOSITY IS STILL FINITE ABOVE $\phi>\phi_\eta$.

For {\it polydisperse} systems, both $P^*$ and $\eta^*$ behave as 
$(\phi_J - \phi)^{-2}$ and $(\phi_J - \phi)^{-3}$ near the jamming 
transition point, $\phi_J > \phi_\eta$, as predicted in Refs.~\citen{Otsuki:PTP,Otsuki:PRE}.
However, as the restitution coefficient $e$ approaches unity,
the scaling regime becomes narrower,
and the exponents for the divergence of $P^*$ and $\eta^*$ 
approach values close to $-1$ in the almost elastic case.

%XXX TODO CONTINUE HERE
From these results, we conclude that the predictions for the 
inelastic soft-disk systems in Refs.~\citen{Otsuki:PTP,Otsuki:PRE}
are applicable to the inelastic near-rigid disk systems below the jamming 
transition point, 
but the prediction cannot be used for almost elastic rigid-disk systems.
It seems that $\tau_{c\omega}^{*}$ and
$\tau_\omega^*$ are the only two possible candidates 
to characterize the criterion of this crossover.
%XXX
%among $\tau_E^*$, $\tau_{cE}^*$, $\tau_\omega^*$, and $\tau_{c\omega}^{*}$.
%However, the validity of the criterion is not sufficient.
%This will be our future work.
In other words, the energy dissipation rate and the shear rate
set the two competing time-scales that define the dimensionless
number $\tau_\omega^*$.  For $\tau_\omega^* \ll 0.01$ the near-rigid, dissipative scaling 
regime occurs, while for $\tau_\omega^* \gg 0.01$ the rigid, elastic
scaling regime is realized.
%XXX PLEASE CHECK  -- should the $\gg$ and $\ll$ be as they are now?

%The range of validity of either scaling relation seems to
%depend on {\em two} dimensionless ratios of the three
%time scales in the system, on the ``softness'' parameter
%$\tau^*_c$ and on the ratio of contact duration to time
%between interactions, $\tau_{cE}^*$. 
%{\bf Feel free to change this
%paragraph if you have a better explanation.}
%However, the complex interplay between contact-duration (softness), 
%local shear-time, and inverse frequency of interactions, is not
%completely sorted out yet.
%{\bf 
%Michio : I think that this sentence should be rewritten by using 
%$\tau_\omega^*$.
%Stefan: I tried but did not succeed. :-(  Let us leave it as is,
%and try to sort it out before we re-submit again.
%Please consider my new part about $\tau_s^*$ ...
%}

%In order to support our current conclusions, data of $\eta^*$
%from polydisperse, elastic, rigid-disk systems are needed to be 
%compared directly with the data presented in this paper.
%However, it is difficult to perform the simulation of hard-disk systems
%in the vicinity of the jamming transition.
%This will be our future work.

In three-dimensional sheared inelastic soft-sphere systems 
\cite{Otsuki:PTP,Otsuki:PRE},
even in {\it monodisperse} cases, there is no indication of the 
strong ordering transition,
and the scaling given in Eqs.\ \eqref{scaling:P} and \eqref{scaling:eta}
seems to be valid. However, a direct comparison of near-rigid sphere
with rigid sphere simulations in the spirit of the present study
is unavailable to our knowledge.

We restricted our interest to frictionless particles.
When the particles have friction, the scaling relations 
for the divergence of the viscosity and the pressure
may be different, as will be discussed elsewhere.
Furthermore, the very soft or
high shear rate regime also needs further attention in both 2D and 3D.

\section*{Acknowledgements}
This work was supported by the Grant-in-Aid for scientific
research from the Ministry of Education, Culture, Sports,
Science and Technology (MEXT) of Japan 
(Nos.~21015016, 21540384, and 21540388),
by the Global COE Program
``The Next Generation of Physics, Spun from Universality
and Emergence'' from MEXT of Japan,
and in part by the Yukawa International Program for
Quark-Hadron Sciences at Yukawa Institute for Theoretical
Physics, Kyoto University.
The numerical calculations were carried out on Altix3700 BX2 at YITP in Kyoto University.
SL acknowledges the hospitality at YITP in Kyoto, 
and support from the Stichting voor Fundamenteel Onderzoek der Materie (FOM),
financially supported by the Nederlandse Organisatie voor Wetenschappelijk
Onderzoek (NWO).

\appendix
\section{The relation between $Z$ and $\tau_{cE}^*$} %Empty argument \section{} yields `Appendix'. 
\label{Z:app}
%
%YYY please check and confirm my changes to the appendix - nothing serious, just to be sure
%
% I am not sure that I got the reasoning for $i=1$ and $j=1$ correct.
% This is connected to the ensemble average, and there I do not see why
% there appears $N-1$. Why not $N$?  And why is there a prefactor anyway,
% that should be divided out in the definition of the ensemble average?
% Please let us iterate once more. Thanks!
%

In this appendix, we derive the relation between $Z$ and $\tau_{cE}^*$ as
\begin{equation}
Z \simeq t_c t_E^{-1} = \tau_{cE}^*,
\label{Z}
\end{equation}
which corresponds to the difference between counting contacts vs. counting
of collisions in the simulations. (Note that counting contacts is not
possible for rigid disks, since the probability to observe a $t_c=0$
contact at any given snapshot in time is zero.)

%The coordination number $Z=\sum_i Z_i/N$ in Eq.\ (\ref{eq:defZ}) 
%is defined as the average number of contacts
%per particle $i$, which has
%\begin{equation}
%Z_i = \sum_{j\neq i} \langle \Theta (\sigma_{ij} - r_{ij}) \rangle
%\label{Z:def}
%\end{equation}
%contacts, where the brackets $\langle \ldots \rangle$ denote an ensemble average.  
%Since the ensemble average is independent of $j$, without loss of generality,
Since the ensemble average in Eq.\ (\ref{eq:defZ}) 
is independent of $i$ and $j$, without loss of generality,
one can set $i=1$ and $j=2$, and obtains
\begin{equation}
\sum_i \sum _{j\neq i} \langle \Theta(\sigma_{ij} - r_{ij}) \rangle
 = N(N-1) \langle \Theta (\sigma_{12} - r_{12}) \rangle.
\end{equation}
Substituting this equation into Eq.\ (\ref{eq:defZ}),
we obtain
\begin{equation}
Z = (N-1) \langle \Theta (\sigma_{12} - r_{12}) \rangle.
\label{Z:eq}
\end{equation}
%\begin{equation}
%\sum_i Z_i = (N-1) \langle \Theta (\sigma_{12} - r_{12}) \rangle.
%\label{Z:eq}
%\end{equation}

On the other hand, $t_E^{-1}$ is defined as the frequency of collisions per 
particle:
\begin{equation}
t_E^{-1}  =  \sum_{j\neq i} \left \{ \lim_{T\to \infty}
\frac{1}{T}n_{c,ij}(T) \right \},
\label{tE:def}
\end{equation}
where $n_{c,ij}(t)$ is the number of the collisions between 
grains $i$ and $j$ until time $T$.
Since $\lim_{T\to \infty} n_{c,ij}(T) /T$
is independent of $j$, like above, we obtain
\begin{equation}
t_E^{-1}  =  (N-1) \left \{ \lim_{T\to \infty} \frac{1}{T}n_{c,12}(T) \right \}
\label{tE:eq}
\end{equation}

In order to derive Eq.\ \eqref{Z},
the ensemble average in Eq.\ \eqref{Z:eq} is replaced by
the time average as
\begin{equation}
Z  = (N-1) \lim_{T\to \infty} \frac{1}{T}
\int_0^{T} dt \ \Theta(\sigma_{12} - r_{12}(t)),
\label{Z:time}
\end{equation}
where $r_{ij}(t)$ is the distance between grains $i$ and $j$ at time $t$.
Since $\Theta(\sigma_{12} - r_{12}(t)) = 1$ for the duration $t_c$ after a
collision begins, the integral in Eq.\ \eqref{Z:time} is estimated as
$n_{c, 12}(T) \ t_c$, which yields
\begin{eqnarray}
Z  & = &
(N-1) \lim_{T\to \infty} \frac{1}{T}
\{ n_{c,12}(T)  t_c \} \nonumber \\
& = & \left [ (N-1) \left \{ \lim_{T\to \infty} \frac{1}{T}
n_{c,12}(T) \right \} \right ]  t_c.
\label{Z:final}
\end{eqnarray}
Finally, substituting  Eq.\ \eqref{tE:eq} into this equation,
gives Eq.\ \eqref{Z} so that we can apply Eq.\ \eqref{eq:defZ}.

%\section{Second Appendix}

\end{document}